\documentclass[fleqn,usenatbib]{mnras}

\usepackage{newtxtext,newtxmath}

\usepackage[T1]{fontenc}
\usepackage{ae,aecompl}

\usepackage{graphicx}	
\usepackage{amsmath}	
\usepackage{longtable}
\usepackage[utf8]{inputenc}

\title[Type Iax SN 2019muj]{SN 2019muj -- a well-observed Type Iax supernova that bridges the luminosity gap of the class}

\author[B. Barna et al.]{Barnab\'as Barna,$^{1,2}$\thanks{E-mail: bbarna@titan.physx.u-szeged.hu}
Tam\'as Szalai,$^{2,3}$
Saurabh W. Jha,$^{4}$
Yssavo Camacho-Neves,$^{4}$ \newauthor
Lindsey Kwok,$^{4}$
Ryan J. Foley,$^{5}$
Charles D. Kilpatrick,$^{5}$
David A. Coulter,$^{5}$ \newauthor
Georgios Dimitriadis,$^{5}$
Armin Rest,$^{6,7}$ 
C\'esar Rojas-Bravo,$^{5}$ 
Matthew R. Siebert,$^{5}$ \newauthor
Peter J. Brown,$^{8,9}$
Jamison Burke,$^{10,11}$ 
Estefania Padilla Gonzalez,$^{10,11}$ \newauthor
Daichi Hiramatsu,$^{10,11}$
D. Andrew Howell,$^{10,11}$ 
Curtis McCully,$^{10}$ 
Craig Pellegrino,$^{10,11}$ \newauthor
Matthew Dobson,$^{12}$ 
Stephen J. Smartt,$^{12}$
Jonathan J. Swift,$^{13}$ 
Holland Stacey,$^{13}$ \newauthor
Mohammed Rahman,$^{13}$ 
David J. Sand,$^{14}$
Jennifer Andrews,$^{14}$
Samuel Wyatt,$^{14}$ \newauthor
Eric Y. Hsiao,$^{15}$ 
Joseph P. Anderson,$^{16}$
Ting-Wan Chen,$^{17}$
Massimo Della Valle,$^{18}$ \newauthor
Llu\'is Galbany,$^{19}$ 
Mariusz Gromadzki,$^{20}$ 
Cosimo Inserra,$^{21}$
Joe Lyman,$^{22}$ \newauthor
Mark Magee,$^{23}$ 
Kate Maguire,$^{23}$
Tom\'as E. M\"uller-Bravo,$^{24}$
Matt Nicholl,$^{25,26}$ \newauthor
Shubham Srivastav,$^{12}$
Steven C. Williams$^{27,28}$ 
\\
$^{1}$Astronomical Institute, Czech Academy of Science, Bocni II 1401, 141 00 Prague, Czech Republic\\ 
$^{2}$Department of Optics and Quantum Electronics, University of Szeged, D\'om t\'er 9, Szeged, 6720 Hungary\\
$^{3}$Konkoly Observatory, Research Centre for Astronomy and Earth Sciences, H-1121 Budapest, Konkoly Thege Mikl\'os út 15-17, Hungary\\
$^{4}$Department of Physics and Astronomy, Rutgers, the State University of New Jersey, 136 Frelinghuysen Road, Piscataway, NJ 08854, USA\\
$^{5}$Department of Astronomy and Astrophysics, University of California, Santa Cruz, CA 95064, USA\\
$^{6}$Space Telescope Science Institute, 3700 San Martin Drive, Baltimore, MD 21218, USA\\
$^{7}$Department of Physics and Astronomy, The Johns Hopkins University, Baltimore, MD 21218, USA\\
$^{8}$Department of Physics and Astronomy, Texas A$\&$M University, 4242 TAMU, College Station, TX 77843, USA\\
$^{9}$George P. and Cynthia Woods Mitchell Institute for Fundamental Physics \& Astronomy, USA\\
$^{10}$Las Cumbres Observatory, 6740 Cortona Drive, Suite 102, Goleta, CA 93117-5575, USA\\
$^{11}$Department of Physics, University of California, Santa Barbara, CA 93106-9530, USA\\
$^{12}$Astrophysics Research Centre, School of Mathematics and Physics, Queen's University Belfast, Belfast, BT7 1NN, UK\\
$^{13}$Thacher Observatory, Thacher School, 5025 Thacher Rd. Ojai, CA 93023, USA\\
$^{14}$Steward Observatory, University of Arizona, 933 North Cherry Avenue, Tucson, AZ 85721-0065, US\\
$^{15}$Department of Physics, Florida State University, 77 Chieftan Way, Tallahassee, FL 32306, USA\\
$^{16}$European Southern Observatory, Alonso de C\'ordova 3107, Casilla 19 Santiago, Chile\\
$^{17}$The Oskar Klein Centre, Department of Astronomy, Stockholm University, AlbaNova, SE-10691 Stockholm, Sweden\\
$^{18}$INAF- Capodimonte, Naples, Salita Moiariello 16, 80131, Naples, Italy\\
$^{19}$Departamento de F\'isica Te\'orica y del Cosmos, Universidad de Granada, E-18071 Granada, Spain\\
$^{20}$Astronomical Observatory, University of Warsaw, Al. Ujazdowskie 4, 00-478 Warszawa, Poland\\
$^{21}$School of Physics \& Astronomy, Cardiff University, Queens Buildings, The Parade, Cardiff, CF24 3AA, UK\\
$^{22}$Department of Physics, University of Warwick, Coventry CV4 7AL, UK\\
$^{23}$School of Physics, Trinity College Dublin, The University of Dublin, Dublin 2, Ireland\\
$^{24}$School of Physics and Astronomy, University of Southampton, Southampton, Hampshire, SO17 1BJ, UK\\
$^{25}$Birmingham Institute for Gravitational Wave Astronomy and School of Physics and Astronomy, University of Birmingham, \\Birmingham B15 2TT, UK \\
$^{26}$Institute for Astronomy, University of Edinburgh, Royal Observatory, Blackford Hill, EH9 3HJ, UK\\
$^{27}$Finnish Centre for Astronomy with ESO (FINCA), Quantum, Vesilinnantie 5, University of Turku, 20014 Turku, Finland\\
$^{28}$Department of Physics and Astronomy, University of Turku, 20014 Turku, Finland\\
}

\date{Accepted XXX. Received YYY; in original form ZZZ}

\pubyear{2020}

\begin{document}
\label{firstpage}
\pagerange{\pageref{firstpage}--\pageref{lastpage}}
\maketitle

\clearpage

\begin{abstract}

We present early-time ($t < +50$ days) observations of SN 2019muj (= ASASSN-19tr), one of the best-observed members of the peculiar SN Iax class. Ultraviolet and optical photometric and optical and near-infrared spectroscopic follow-up started from $\sim$5 days before maximum light ($t_{\rm max}(B)$ on $58\,707.8$ MJD) and covers the photospheric phase. The early observations allow us to estimate the physical properties of the ejecta and characterize the possible divergence from a uniform chemical abundance structure. The estimated bolometric light curve peaks at 1.05 $\times$ 10$^{42}$ erg s$^{-1}$ and indicates that only 0.031 $M_\odot$ of $^{56}$Ni was produced, making SN 2019muj a moderate luminosity object in the Iax class with peak absolute magnitude of $M_\rmn{V} = -16.4$ mag. The estimated date of explosion is $t_0$ = $58\,698.2$ MJD and implies a short rise time of $t_{\rm rise}$ = 9.6 days in $B$-band. We fit of the spectroscopic data by synthetic spectra, calculated via the radiative transfer code {\tt TARDIS}. Adopting the partially stratified abundance template based on brighter SNe Iax provides a good match with SN 2019muj. However, without earlier spectra, the need for stratification cannot be stated in most of the elements, except carbon, which is allowed to appear in the outer layers only. SN 2019muj provides a unique opportunity to link extremely low-luminosity SNe Iax to well-studied, brighter SNe Iax.

\end{abstract}




\section{Introduction}
\label{sec:introduction}

Although most normal Type Ia supernovae (SNe Ia) form a homogeneous class \citep[often referred to as `normal' or `Branch-normal' SNe,][]{Branch06}, a growing number of peculiar thermonuclear explosions are being discovered \citep{Taubenberger17}. These objects are also assumed to originate from a carbon-oxygen white dwarf (C/O WD), but do not follow the correlation between the shape of their light curve (LC) and their peak luminosity \citep{Phillips93} controlled by the synthesized amount of $^{56}$Ni \citep{Arnett82}. These peculiar thermonuclear SNe also show unusual observables compared to those of normal SNe Ia.

One of these subclasses is the group of Type Iax SNe \citep[SNe Iax;][]{Foley13}, or, as named after the first discovered object, `2002cx-like' SNe \citep{Li03,Jha06}. \citet{Jha17} present a collection of $\sim$60 SNe Iax, making the group probably the most numerous of the Ia subclasses. However, the exact rate of SNe Iax is highly uncertain as \cite{Foley13} found it to be 31$^{+17}_{-13}$\% of the normal SN Ia rate based on a volume-limited sample, consistent with results of \cite{Miller17}. The peak absolute magnitudes of SN Iax cover a wide range between -14.0 -- -18.4 mag, from the extremely faint SNe 2008ha \citep{Foley09,Valenti09} and 2019gsc \citep{Srivastav20,Tomasella20} to the nearly normal Ia-bright SNe 2011ay \citep{Szalai15,Barna17} and 2012Z \citep{Stritzinger15}. The distribution is probably dominated by the faint objects and the number of the brighter (M$_\rmn{V} < -17.5$ mag) ones is only 2-20\% of the total SNe Iax population \citep{Li11,Graur17}. The expansion velocities also have a very diverse nature, showing a photospheric velocity ($v_\rmn{phot}$) of 2,000--9,000 km s$^{-1}$ at the moment of maximum light, which is significantly lower than that of SNe Ia \citep[typically $> 10\,000$ km s$^{-1}$,][]{Silverman15}. A general correlation between the peak luminosity and the expansion velocity (practically, $v_\rmn{phot}$ at the moment of maximum light) of SNe Iax has been proposed by \citet{McClelland10}. However, the level of such correlation is under debate, mainly because of reported outliers like SNe 2009ku \citep{Narayan11} and 2014ck \citep{Tomasella16}.

The spectral analysis of SNe Iax explored a similar set of lines in the optical regime as in the case of normal SNe Ia \citep{Phillips07, Foley10a}. However, the early phases are dominated mainly by the features of iron-group elements (IGEs), especially the absorption features of Fe \begin{small}II\end{small} and Fe \begin{small}III\end{small}. While Si \begin{small}II\end{small} and Ca \begin{small}II\end{small} are always present, their characteristic lines are not optically thick, and high-velocity features \citep{Silverman15} have not been observed in SNe Iax spectra.

\cite{McCully14} reported the discovery of a luminous blue point source coincident with the location of SN 2012Z in the pre-explosion HST images. The observed source was similar to a helium nova system, leading to its interpretation as the donor star for the exploding WD. SN 2012Z could be the first ever thermonuclear SN with a detected progenitor system, and thus linked to the single degenerate (SD) scenario by direct observational evidence. The model of SNe Iax originating from WD-He star systems is supported by the generally young ages of SNe Iax environments \citep{Lyman18,Takaro20}, though at least one SN Iax exploded in an elliptical galaxy \citep{Foley10b}. The detection of helium is also reported in the spectra of SNe 2004cs and 2007J \citep{Foley13,Jacobson19,Magee19}, but there is a debate on the real nature of these two objects \cite[see ][]{White15,Foley16}. Moreover, the apparent lack of helium in the majority of known SNe Iax \citep{Magee19} shows the necessity of further investigation of this point.

However, one of the most critical questions regarding the Iax class is whether these SNe with widely varying luminosities could originate from the same explosion scenario? Among the existing theories, the most popular scenario \citep{Jha17} in the literature is the pure deflagration of a Chandrasekhar-mass C/O WD leaving a bound remnant behind, which is supported by the predictions of hydrodynamic explosion models \citep{Jordan12b, Kromer13, Fink14, Kromer15}. A modified version of this theory, in which the progenitor is a hybrid CONe WD, has been also published in several papers to explain the origin of the faintest members of the class \citep[see e.g.][]{Denissenkov15, Kromer15, Bravo16}.

As a further discrepancy compared to normal SN Ia is that SNe Iax never enter into a truly nebular phase like normal SNe Ia, instead showing both P Cygni profiles of permitted lines and emission of forbidden transitions $\sim$200 days after the explosion \citep{Jha06}. A possible explanation of the late-time evolution comes from \cite{Foley14}, who proposed a super-Eddington wind from the bound remnant as the source of the late-time photosphere. The bound remnant was originally predicted by hydrodynamic simulations of pure deflagration of a Chandrasekhar-mass WD \citep{Fink14}. Such a remnant may have been seen in late-time observations of SN 2008ha \citep{Foley14}. Understanding the dual nature of these spectra requires further investigation and observations of extremely late epochs.

The deflagration scenarios that fail to completely unbind the progenitor WD are able to reproduce the extremely diverse luminosities and kinetic energies of the Iax class. On the other hand, the synthetic light curves from the pure deflagration models of \cite{Kromer13} and \cite{Fink14} seem to evolve too fast after their maxima, indicating excessively low optical depths, i.e. insufficient ejecta masses. Nevertheless, it has to be noted that only a small range of initial parameters of deflagration models leaving a bound remnant has been sampled in these studies; therefore, it cannot be excluded that with different initial conditions (ignition geometry, density, composition) larger ejecta masses (for a given $^{56}$Ni mass) are possible in the deflagration scenario.

Apart from the pure deflagration, other single-degenerate (SD) explosion scenarios, which may partially explain the observables of SNe Iax, are the deflagration-to-detonation transition \citep[DDT;][]{Khokhlov91a} scenarios. In classical DDT models, the WD becomes unbound during the deflagration phase; in a variation, the pulsational delayed detonation (PDD) scenario, the WD remains bound at the end of the deflagration phase, then undergoes a pulsation followed by a delayed detonation \citep{Ivanova74,Khokhlov91b,Hoflich95}. The detonation can happen via a sudden energy release in a confined fluid volume; this group of models includes gravitationally confined detonations \citep[GCD, see e.g.][]{Plewa04,Jordan08}, ‘pulsationally assisted’ GCD models \citep{Jordan12a} and pulsating reverse detonation (PRD) models \citep{Bravo06,Bravo09}. From all of these delayed detonation scenarios, only PDD models have been used for direct comparison with the data of an SN Iax \citep[SN 2012Z,][]{Stritzinger15}.

A common feature of all the DDT scenarios is the stratified ejecta structure caused by the mechanism of the transition from deflagration-to-detonation. On the other hand, one of the characteristic ejecta properties of the pure deflagration scenarios, is the well-mixed ejecta, resulting in nearly constant chemical abundances. Indications for the mixed ejecta structure based on the near-infrared light curves have been reported since the discovery of SNe Iax \citep{Li03,Jha17}. As it was shown by \cite{Kasen06}, the \textit{IJHK} light curves provide a diagnostic tool for the mixing of IGEs, which elements drive the formation of a secondary maximum by drastic opacity change due to their recombination \citep[see e.g.][]{Pinto00,Hoflich95,Hoflich17}. Although several parameters (e.g. progenitor metallicity, abundance of IMEs) have also a significant impact on its formation, the existence of the NIR secondary peak is mainly a consequence of the concentration of IGEs in the central region of SN Ia ejecta. \cite{Phillips07} showed, that the NIR LCs of SN 2005hk, where the first and second maxima are indistinguishable, is similar to that of the SN Ia model containing 0.6 $M_\odot$ of $^{56}$Ni in a completely homogenized composition structure \citep{Kasen06}. The authors argued that despite the different model, the lack of a separated secondary peak in the NIR LCs is a direct evidence for mixed ejecta structure of SNe Iax.

At the same time, the pure deflagration producing a bound remnant picture might contradict the bright SN Iax 2012Z \citep{Stritzinger15}, in which the velocity distribution of intermediate mass elements (IMEs) do not indicate significant mixing in the outer ejecta. \cite{Barna18} performed abundance tomography for a small sample of SNe Iax and found a stratified structure as a feasible solution for the outer regions of SNe Iax, however, the impact of the outermost layers are strongly affected by the choice of the density profile.

Despite the fact that the former results in this question are contradictory, constraining the chemical structure could be the key to either confirm the assumption of the pure deflagration scenario or reject it. Note that with the term (pure) deflagration we refer only on the weak deflagration scenarios hereafter, which leave behind a bound remnant \citep{Fink14,Kromer15}, and thus, provide a promising explanation for the SN Iax class.

A promising way to test the theory whether SN Iax share a similar origin (i.e. the pure deflagration of a CO/CONe $M_\rmn{Ch}$ WD) is the investigation of the link between the two extremes of the Iax class, the relatively luminous and extremely faint objects. A continuous distribution of physical and chemical properties, as well as their possible correlations with the peak luminosity, favor the idea that SNe Iax may form a one (or maybe a few) parameter family. So far, these efforts are limited by the lack of well-observed, moderately luminous SNe Iax in between the extremes. The subject of our current study, SN 2019muj, now provides a good opportunity to carry out a detailed analysis of an SN Iax belongs to this luminosity gap.

\begin{figure}
	\includegraphics[width=\columnwidth]{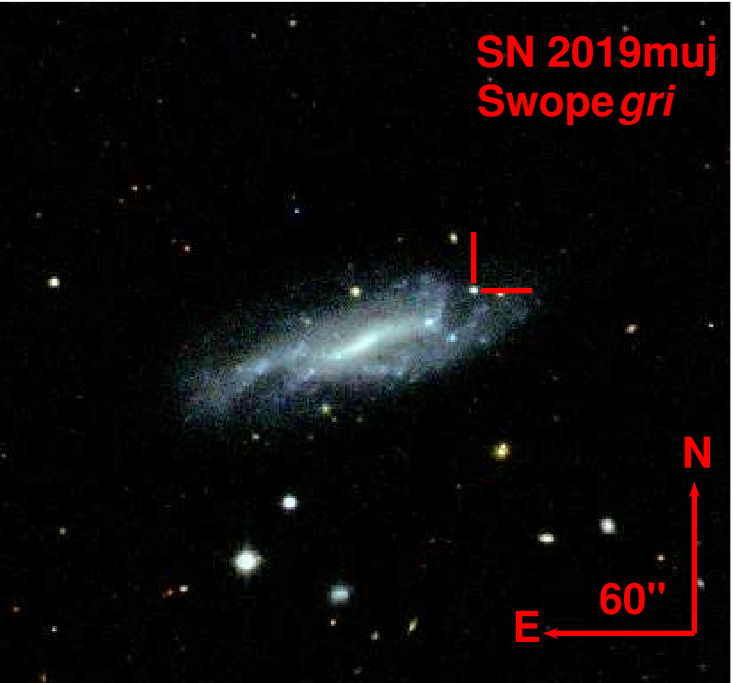}
    \caption{The $gri$ composite image of the field with the VV 525 galaxy and SN 2019muj obtained by Swope telescope at the Las Campanas Observatory.}
    \label{fig:disc_image}
\end{figure}

The paper is structured in the following order. In Section \ref{target}, we introduce SN 2019muj and give an overview of the collected data. The light curves, spectra, and the estimated ultraviolet (UV)-optical spectral energy distributions (SEDs) are shown in Section \ref{discussion}. We summarize our conclusions in Section \ref{conclusions}.

\section{Observations and Data reduction}
\label{target}

\subsection{SN 2019muj}
\label{19muj}

SN 2019muj (ASASSN-19tr) was discovered \citep{Brimacombe19} by the All Sky Automated Survey for SuperNovae \citep[ASAS-SN;][]{Shappee14} on 7 August 2019, UT 09:36 (MJD $58\,702.4$). The location in the sky (R.A. 02:26:18.472, Dec. $-$09:50:09.92, 2000.0) is associated with VV 525, a blue star-forming spiral galaxy (see Fig. \ref{fig:disc_image}) with morphological type of SAB(s)dm: \citep{deVaucouleurs91} at redshift of $z=0.007035$. Considering the low redshift, the K-correction is assumed negligible \citep{Hamuy93,Phillips07}. We adopted $d=34.1 \pm 2.9$ Mpc for the distance of the galaxy from \cite{Tully13,Tully16} (using the lower uncertainty value from the two studies) calculated by the Tully-Fisher method, assuming $H_0 = 70$ km s$^{-1}$ Mpc$^{-1}$. This value is slightly higher than the distance of $d_\rmn{z}=30.2$ Mpc estimated from the measured redshift. Hereafter, we use the former distance for spectral modelling and the calculation of the quasi-bolometric LC.

The SN is in the outskirts of the galaxy at a 48'' projected separation (8 kpc) from its center (see Fig. \ref{fig:disc_image}). Because of the lack of significant Na \begin{small}I\end{small} D line absorption at the redshift of the galaxy, we assume the host galaxy reddening at the SN position as $E(B-V)_\rmn{host} = 0.0$, while the Galactic contribution is adopted as $E(B-V)_\rmn{Gal} = 0.023$ from \cite{Schlafly11}. Moreover, because the SN is well-separated from its host galaxy, image subtraction was not required during the data reduction.

At discovery, the apparent magnitude of SN 2019muj was 17.4 mag in the {\textit g}-band. Based on the first spectrum, obtained at MJD 58702.79, the Superfit classification algorithm \citep{Howell05} found SN 2019muj similar to SN 2005hk around a week before maximum, thus, categorized as a SN Iax \citep{Hiramatsu19}. These properties indicated SN 2019muj to be the brightest SN Iax with pre-maximum discovery since the discovery of SN 2014dt and intensive follow-up campaigns were started by several collaborations and observatories.

\subsection{Photometry}
\label{photometric_obs}

\begin{figure}
	\includegraphics[width=8.5cm]{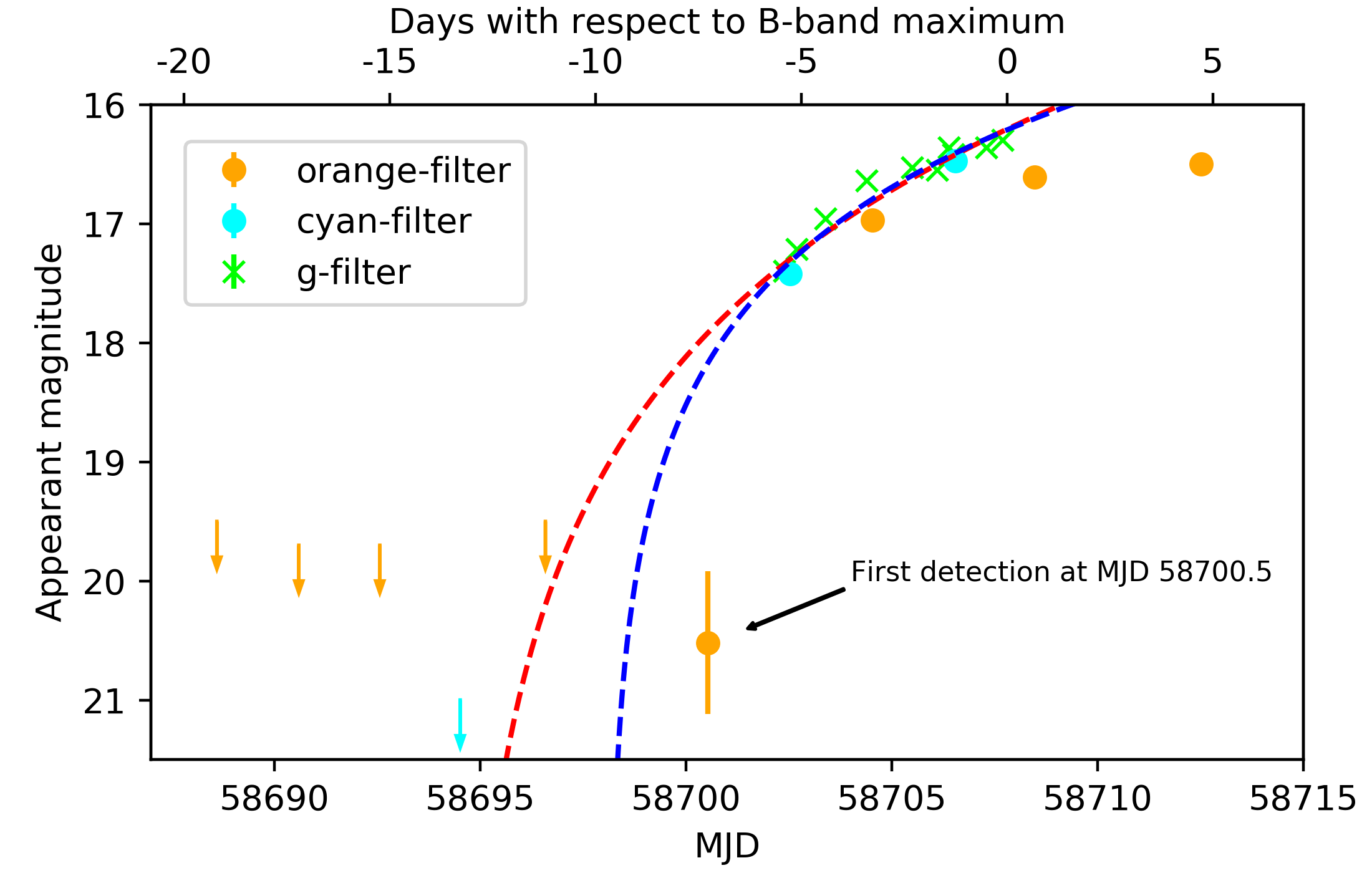}
    \caption{The pre-maximum ATLAS photometry in \textit{cyan-} and \textit{orange}-filters at the location of SN 2019muj. The dashed lines represent the fit of the early {\textit c-}, {\textit o}- and {\textit g-}band LCs (latter observations are also plotted ) with power-law index n=1.3 (blue) and n=2.0 (red). Before MJD $58\,700$, the arrows show the 3$\sigma$ upper limits. Note that the first detection has only 2$\sigma$ significance. For further details see Sec. \ref{photometry}. }
    \label{fig:sn19muj_photometry_ATLAS}
\end{figure}

The field of SN 2019muj was observed by the Asteroid Terrestrial-impact Last Alert System program \citep[ATLAS; ][]{Tonry18} before its discovery, providing a deep ($\sim$21 mag in orange-band) non-detection limit before MJD $58\,700$ (see in Fig. \ref{fig:sn19muj_photometry_ATLAS}). The monitoring in $o$- and $c$-bands continued through the whole observable time range.

The ATLAS data provide a strong constraint on the explosion epoch. ATLAS detected SN2019muj on MJD $58\,702.53$, just a few hours after the ASAS-SN discovery. As with all ATLAS transients, forced photometry was performed \citep[as described in][]{Smith20} around the time of discovery, with the fluxes presented in Table \ref{tab:photometric_data_ATLAS} and plotted in Fig. \ref{fig:sn19muj_photometry_ATLAS}. There is a marginal 1.8$\sigma$ detection on MJD $58\,700.52$, two days before discovery. This can either be interpreted as a detection at $o=20.5\pm0.6$ mag or a 3$\sigma$ upper limit at $o < 20$ mag. Our analysis is not affected by either choice, as discussed further in Sec. \ref{photometry}.

\begin{figure}
	\includegraphics[width=8.5cm]{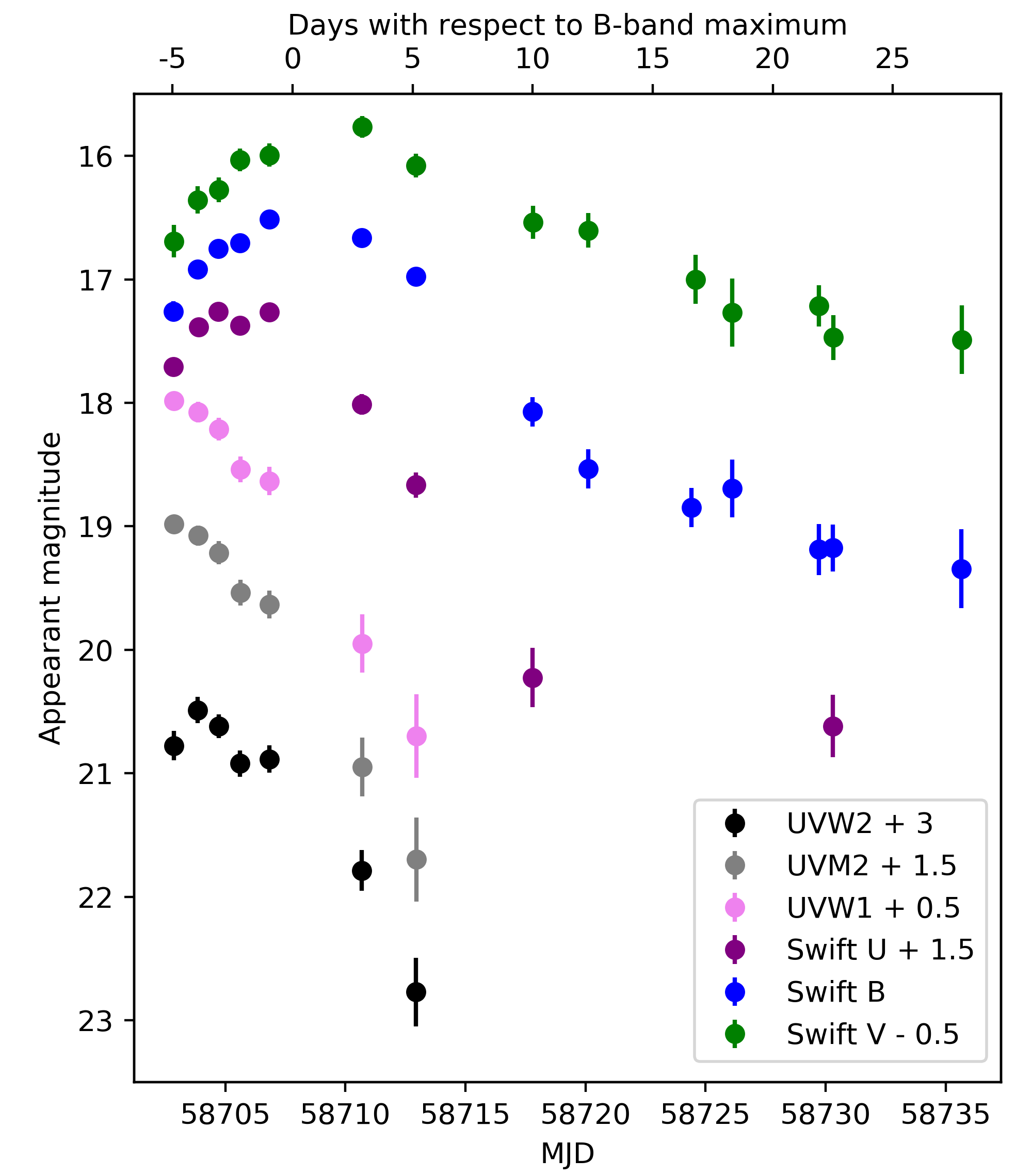}
    \caption{Swift photometry of SN 2019muj. }
    \label{fig:sn19muj_photometry_swift}
\end{figure}

SN 2019muj was also observed with the Neil Gehrels Swift Observatory Ultraviolet/Optical Telescope (UVOT) \citep[hereafter \textit{Swift};][]{Burrows05,Roming05} on 13 epochs between MJD $58\,702.8$ and $58\,735.7$. The \textit{Swift} data were reduced using the pipeline of the Swift Optical Ultraviolet Supernova Archive \citep[SOUSA;][]{Brown14}, without subtraction of the host galaxy flux. \textit{Swift} light curves (see Fig. \ref{fig:sn19muj_photometry_swift}) provide unique UV information in UVW2, UVM2 and UVW1 filters and supplementary information to the ground-based data in $UBV$ (see Table \ref{tab:photometric_data_Swift}). Note that flux conversion factors are spectrum dependent and the UVW2 and UVW1 filters are heavily contaminated by the optical fluxes \citep{Brown10,Brown16}. This red leak may cause high uncertainties for the magnitudes of the wide bands.

Ground-based photometric follow-up was obtained with the Sinistro cameras on Las Cumbres Observatory  network (LCO) of 1-m telescopes at Sutherland (South Africa), CTIO (Chile), Siding Spring (Australia), and McDonald \citep[USA;][]{Brown13}, through the Global Supernova Project (GSP). Data were reduced using {\tt lcogtsnpipe} \citep{Valenti16}, a PyRAF-based photometric reduction pipeline, by performing PSF-fitting photometry. Images in Landolt filters were calibrated to Vega magnitudes, and zeropoints were calculated nightly from Landolt standard fields taken on the same night by the telescope, the corresponding zeropoints and color terms are also calculated in the natural system. Since the target is in the Sloan field, zeropoints for images in Sloan filters were calculated by using Sloan AB magnitudes of stars in the same field as the object. The estimated uncertainties take into account zero points noise, Poisson noise, and the read out noise.

We imaged 2019muj from MJD 58\,702.5 to 58\,908.0 with the Direct 4k$\times$4k imager on the Swope 1-m telescope at Las Campanas Observatory, Chile. We performed observations in Sloan $ugri$ filters and Johnson \textit{BV} filters. All bias-subtraction, flat-fielding, image stitching, registration, and photometric calibration were performed using {\tt photpipe} \citep{Rest05} as described in \citet{Kilpatrick18}. No host-galaxy subtraction was performed, which would cause no significant differences, as SN 2019muj is appeared at the edge of the galaxy. We calibrated our photometry in the Swope natural system \citep{Krisciunas17} using photometry of stars in the same field as 2019muj from the Pan-STARRS DR1 catalog \citep{Flewelling16} and transformed using the Supercal method \citep{Scolnic15}. Final photometry was obtained using {\tt DoPhot} \citep{Schechter93} on the reduced images. For better comparison, the {\textit BV} magnitudes are then transformed to the Vega magnitudes \citep{Blanton07}. The uncertainties are computed by combining the statistical uncertainties in the photometry, and the systematic uncertainties due to the calibration and transformations used.

Additional optical imaging was collected by Thacher Observatory \citep{Swift18} in Ojai, CA from 9 August 2019 to 31 August 2019.  The observations were taken in Johnson $V$ and Sloan $griz$ filters.  All reductions were performed in {\tt photpipe} \citep{Rest05} following the same procedures as the Swope data but with photometric calibration in the PS1 system for $griz$ and using PS1 magnitudes transformed to Johnson $V$ band using the equations in \citet{Jester05}. In addition, the Thacher telescope camera is non-cryogenic, and so we performed dark current subtractions using $60$~second dark frames obtained in the same instrumental configuration and on the same night as our science frames.

\begin{figure}
	\includegraphics[width=9.0cm]{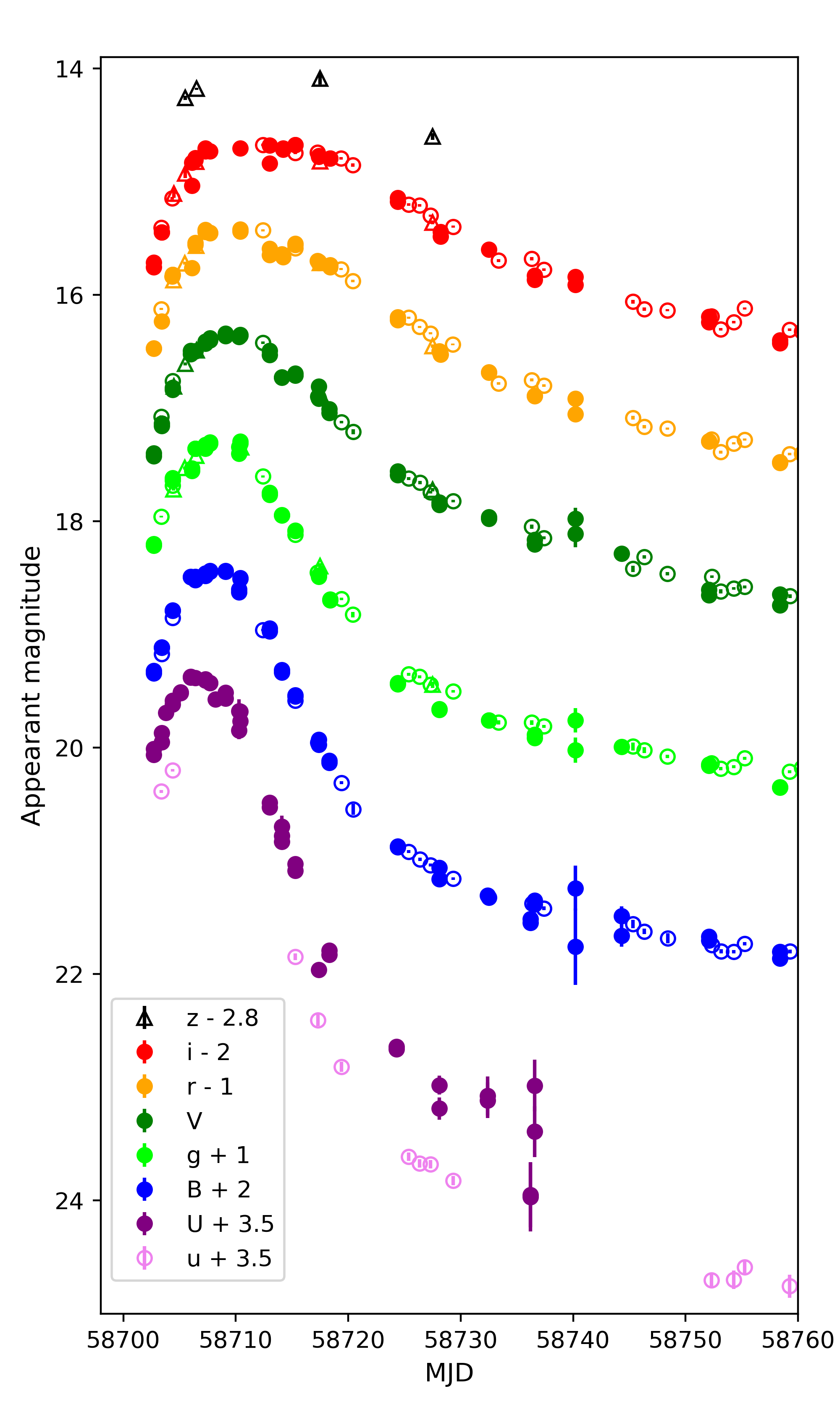}
    \caption{Ground-based photometry of SN 2019muj. The filled circles, open circles, and triangles represent the observations of LCO, Swope and Thacher observatories, respectively. }
    \label{fig:sn19muj_photometry} 
\end{figure}

\begin{figure*}
	\includegraphics[width=14.0cm]{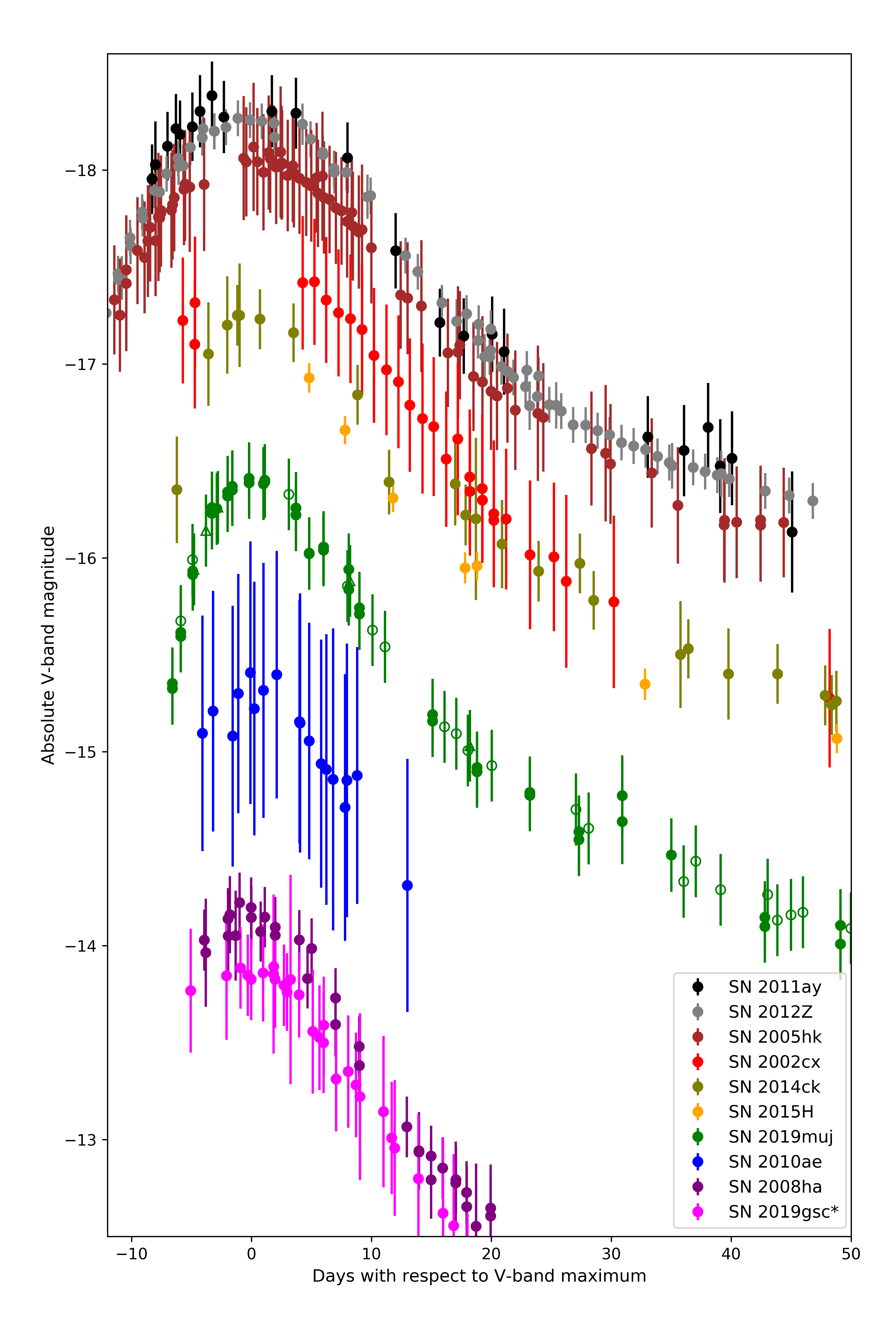}
    \caption{Comparisons of absolute $V$-band light curves of SNe Iax. References to the listed SNe Iax can be found in \citet{Szalai15} for SN 2011ay, \citet{Yamanaka15} and \citet{Stritzinger15} for SN 2012Z, \citet{Phillips07} for SN 2005hk, \citet{Li03} for SN 2002cx, \citet{Tomasella16} for SN 2014ck, \citet{Magee16} for SN 2015H, \citet{Stritzinger14} for SNe 2008ha and 2010ae. Note that in the case of SN 2019gsc \citep{Srivastav20,Tomasella20}, the $V$-band LC is estimated from $g$- and $r$-band magnitudes adopting the transformation presented in \citet{Tonry12}. }
    \label{fig:sn19muj_photometry_v}
\end{figure*}

The ground-based photometric data including the $BVR$ and $ugriz$ magnitudes can be found in Tables \ref{tab:photometric_data_UBV} and \ref{tab:photometric_data_ugriz} respectively, while the LCs are shown in Fig. \ref{fig:sn19muj_photometry}. We compare the absolute $V$-band LC of SN 2019muj to a selection of other SNe Iax that populate the observed range of absolute magnitudes of this class in Fig. \ref{fig:sn19muj_photometry_v}.

\subsection{Spectroscopy}
\label{spectroscopic_obs}

Optical spectroscopy of SN~2019muj was obtained by the 9.2-m Southern African Large Telescope (SALT) with the Robert Stobie Spectrograph \citep[RSS;][]{Smith06} through Rutgers University program 2019-1-MLT-004 (PI: SWJ). The observations were made with the PG0900 grating and 1.5'' wide longslit with a typical spectral resolution $R = \lambda / \Delta  \lambda \approx 1000$. Exposures were taken in four grating tilt positions to cover the optical spectrum from 3500 to 9300 \AA. The data were reduced using a custom pipeline based on standard Pyraf spectral reduction routines and the PySALT package \citep{Crawford10}.

During the first 50 days after explosion, spectra were also obtained five times with the ESO Faint Object Spectrograph and Camera version 2 \citep[EFOSC2,][]{Buzzoni84} at the 3.6-m New Technology Telescope (NTT) of European Southern Observatory as part of the \href{http://www.pessto.org/}{ePESSTO+} collaboration \citep{Smartt15}. These epochs were observed with grisms \#11 (covering 3300-7500 \r{A} and \#16 (6000-9900 \r{A}), or \#13 (3500-9300 \r{A}) (see Table \ref{tab:spectroscopic_data}). The data reduction of the NTT spectra was performed using the PESSTO pipeline\footnote{\url{https://github.com/svalenti/pessto}} \citep{Smartt15}.

LCO optical spectra were taken with the FLOYDS spectrographs mounted on the 2-m Faulkes Telescope North and South at Haleakala (USA) and Siding Spring (Australia), respectively, through the Global Supernova Project. A 2'' slit was placed on the target at the parallactic angle \citep{Filippenko82}. One-dimensional spectra were extracted, reduced, and calibrated following standard procedures using the FLOYDS pipeline \footnote{\url{https://github.com/svalenti/FLOYDS_pipeline}} \citep{Valenti14}. 

Observations using the Kast spectrograph on the Shane-3m telescope of the Lick Observatory \citep{Miller93} were made using the 2''-wide slit, the 452/3306 grism (blue side), the 300/7500 grating (red side) and the D57 dichroic. These data were reduced using a custom {\tt PyRAF}-based KAST pipeline\footnote{\url{https://github.com/msiebert1/UCSC_spectral_pipeline}}, which accounts for bias-subtracting, optical flat-fielding, amplifier crosstalk, background and sky subtraction, flux calibration and telluric corrections using a standard star observed on the same night \citep[procedure described in][]{Silverman20} and at a similar airmass.

Finally, two NIR spectra were taken with the FLAMINGOS-2 spectrograph \citep{Eikenberry08} on Gemini-South.  The data acquisition and reduction are similar to that described in \citet{Sand18}. The JH grism and 0.72'' width longslit were employed, yielding a wavelength range of $\sim$1.0 -- 1.8 $\mu$m and $R \sim1000$.  All data was taken at the parallactic angle with a standard ABBA pattern, and an A0V telluric standard was observed near in airmass and adjacent in time to the science exposures.  The data were reduced in a standard way using the \texttt{F2 PyRAF} package provided by Gemini Observatory, with image detrending, sky subtraction of the AB pairs, spectral extraction, wavelength calibration and spectral combination.  The telluric corrections and simultaneous flux calibrations were determined using the \texttt{XTELLCOR} package \citep{Vacca03}.

\begin{figure}
	\includegraphics[width=8.8cm]{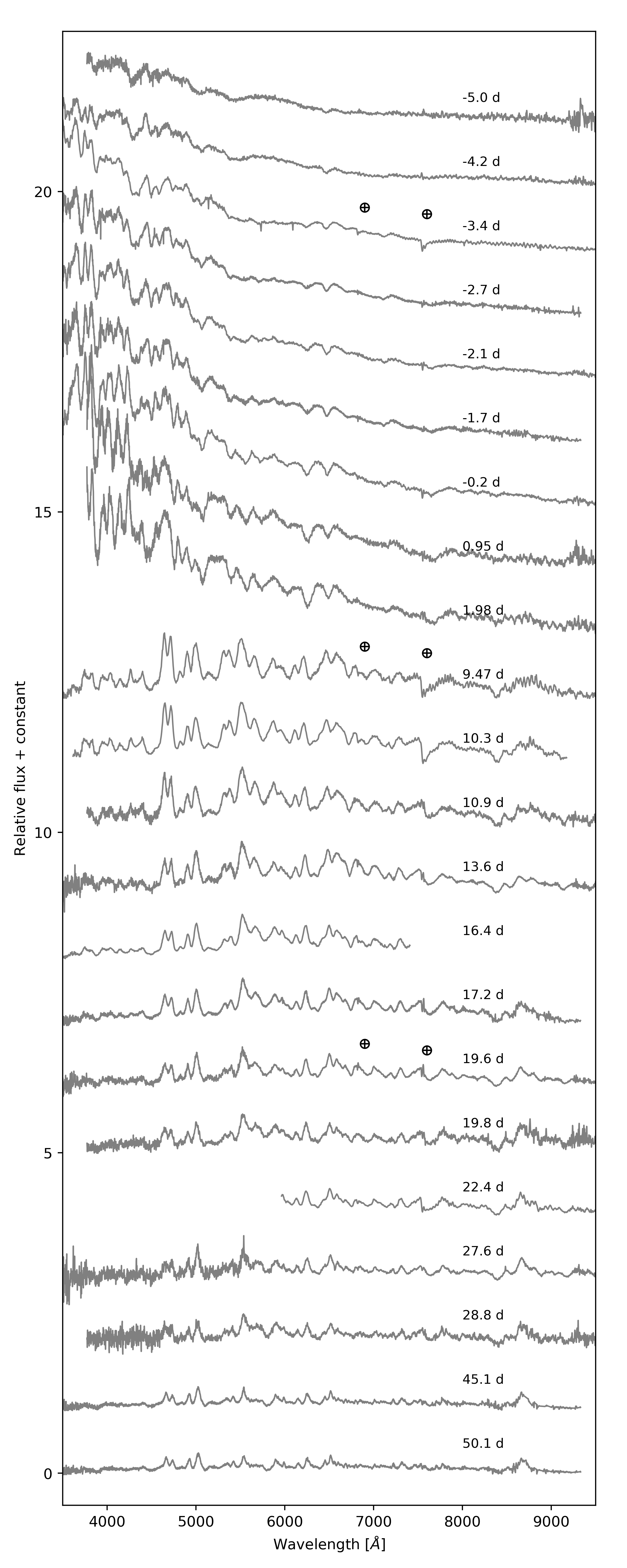}
    \caption{Spectroscopic follow-up of SN 2019muj obtained with SALT/RSS, Lick/Kast, NTT/EFOSC2 and LCO/FLOYDS spectrographs. The epochs show the days with respect to B-maximum. The observation log of spectra can be found in Tab. \ref{tab:spectroscopic_data}. The positions of strong telluric lines are marked with $\oplus$.}
    \label{fig:sn19muj_spectroscopy}
\end{figure}

The optical spectra of SN 2019muj are plotted in Fig. \ref{fig:sn19muj_spectroscopy}, while the log of the spectroscopic observations can be found in Table \ref{tab:spectroscopic_data}.

Our spectroscopic data were not always taken under spectrophotometric conditions; clouds and slit losses can lead to errors in the overall flux normalization. As an example, the pupil illumination of SALT, moreover, changes during the observation, so even relative flux calibration from different grating angles can be difficult. Synthetic photometry of the spectra can differ from the measured photometry by a few tenths of a magnitude. For these reasons, we choose to color match all of our spectroscopic data to match the observed photometry. After this color-matching, the synthetic photometry of the spectra reproduces the photometric data to better than 0.05 mag across optical filters.

All the spectra are scaled and color-matched to the observed broadband photometry. Subsequently, the spectra were corrected for redshift and reddening according to the extinction function of \cite{Fitzpatrick07}. All data will publicly released via WIseREP\footnote{\url{https://wiserep.weizmann.ac.il/}}.

\begin{figure}
	\includegraphics[width=9.0cm]{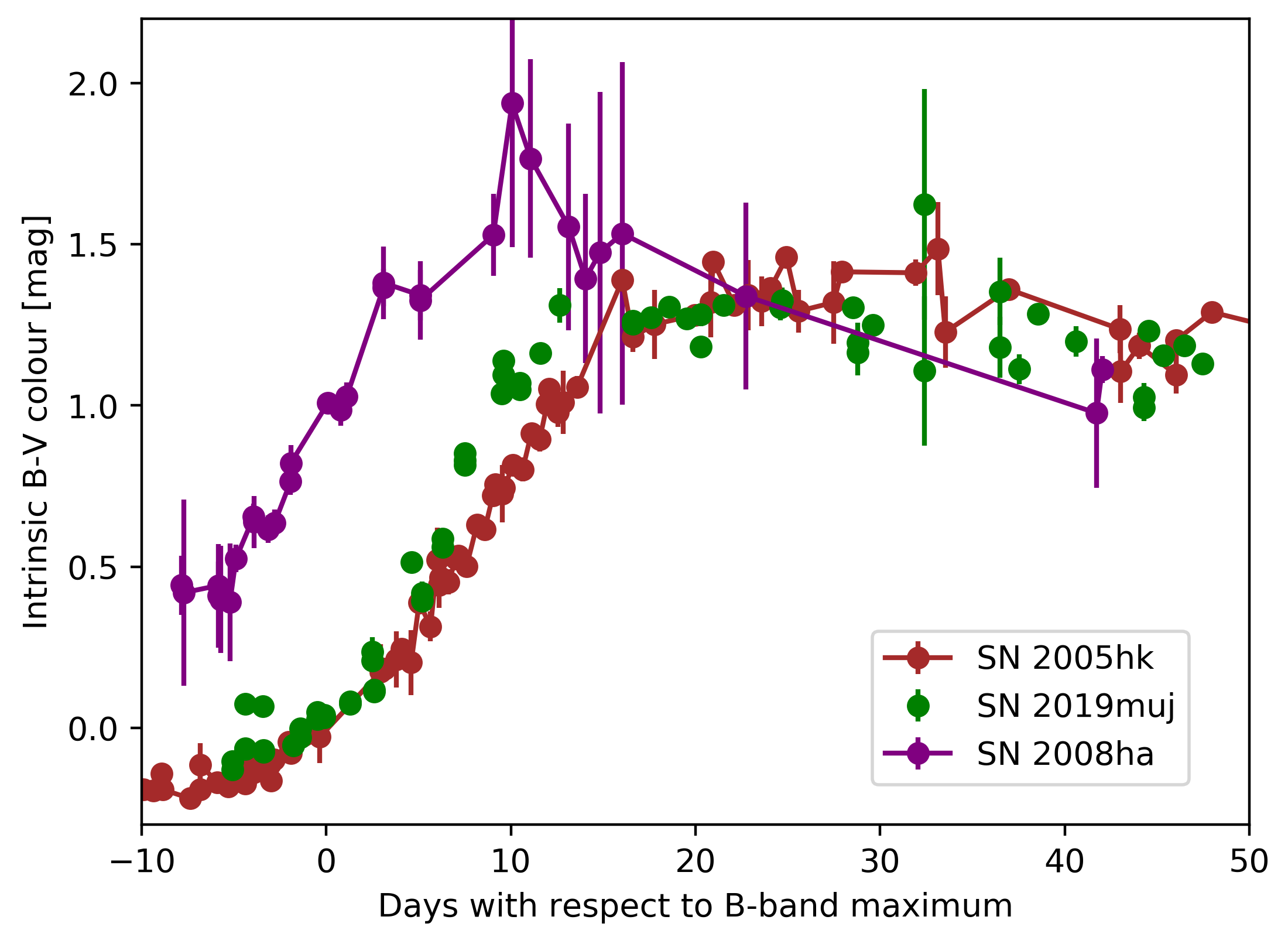}
    \caption{The color evolution of SN 2019muj compared to that of SNe Iax that provide examples of more and less luminous events with respect to SN 2019muj. All magnitudes are corrected for galactic and host extinction with data taken from \citet{Stritzinger14} for SN 2008ha and \citet{Phillips07} for SN 2005hk. }
    \label{fig:color_comparison}
\end{figure}

\begin{figure}
	\includegraphics[width=9.0cm]{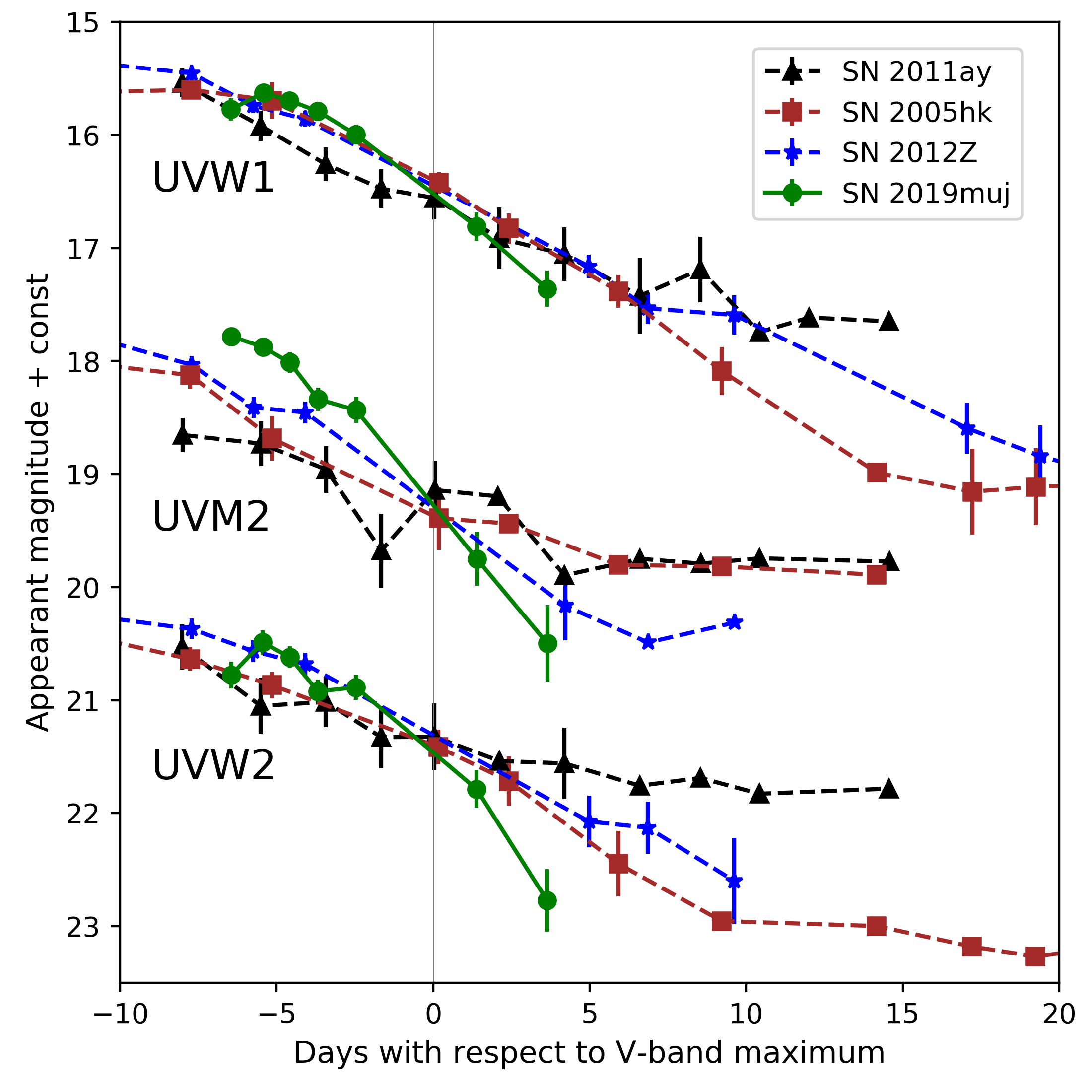}
    \caption{The observed UVW2-, UVM2- and UVW1-band light curves of SNe 2011ay \citep{Szalai15}, 2012Z \citep{Stritzinger15}, 2005hk \citep{Phillips07} and 2019muj. For better comparison, the light curves are shifted along both the horizontal and vertical axes to match at the moment of their V-band maximum.}
    \label{fig:swift_comparison}
\end{figure}

\begin{figure}
	\includegraphics[width=8.2cm]{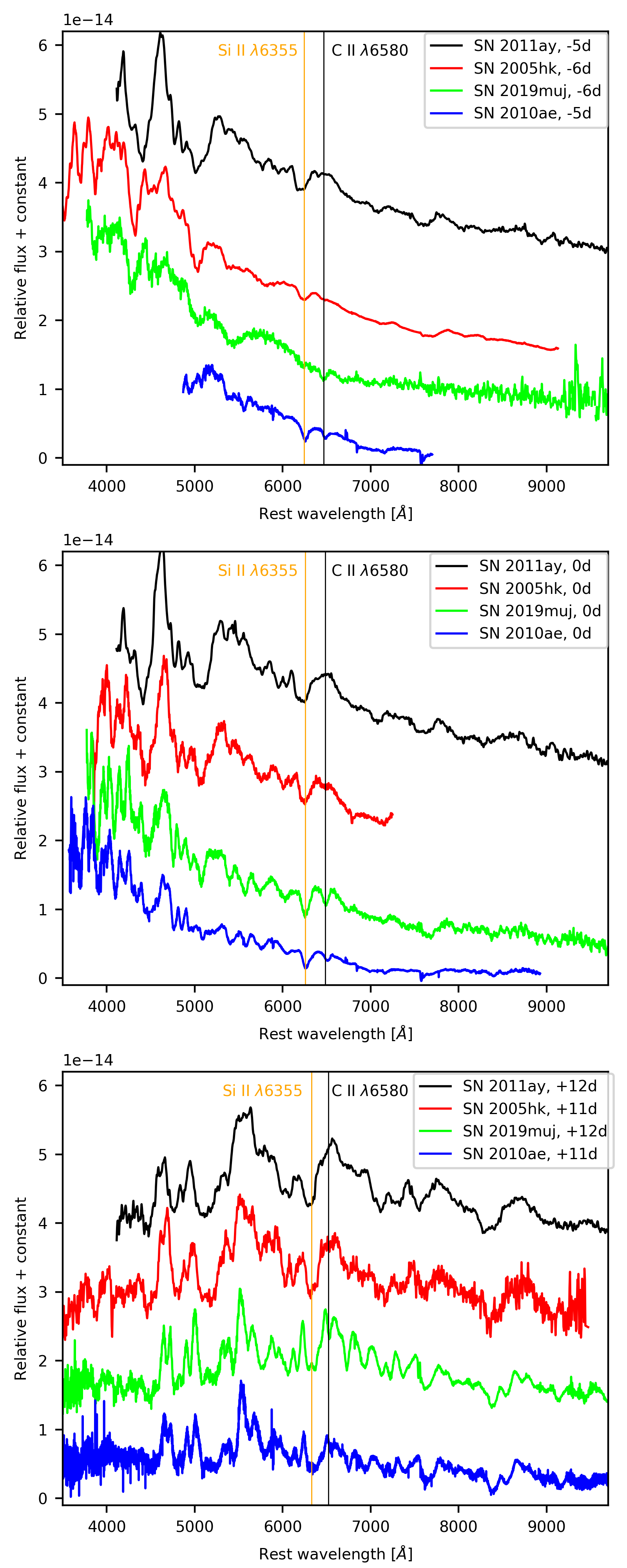}
    \caption{Comparison between spectra of Type Iax SNe 2011ay, 2005hk, 2010ae, and 2019muj obtained at similar epochs. References can be found in \citet{Szalai15}, \citet{Phillips07} and \citet{Stritzinger14}, respectively. The orange and black vertical lines show the absorption minima of the Si II $\lambda6355$ and C II $\lambda$6580 lines in the SN 2019muj spectra. With no visible absorption of the Si II $\lambda6355$, the orange line show its position blueshifted with 5600 km s $^{-1}$ (the $v_\rmn{phot}$ of the first TARDIS model, see below).}
    \label{fig:comparison_iax}
\end{figure}

\begin{figure*}
	\includegraphics[width=14cm]{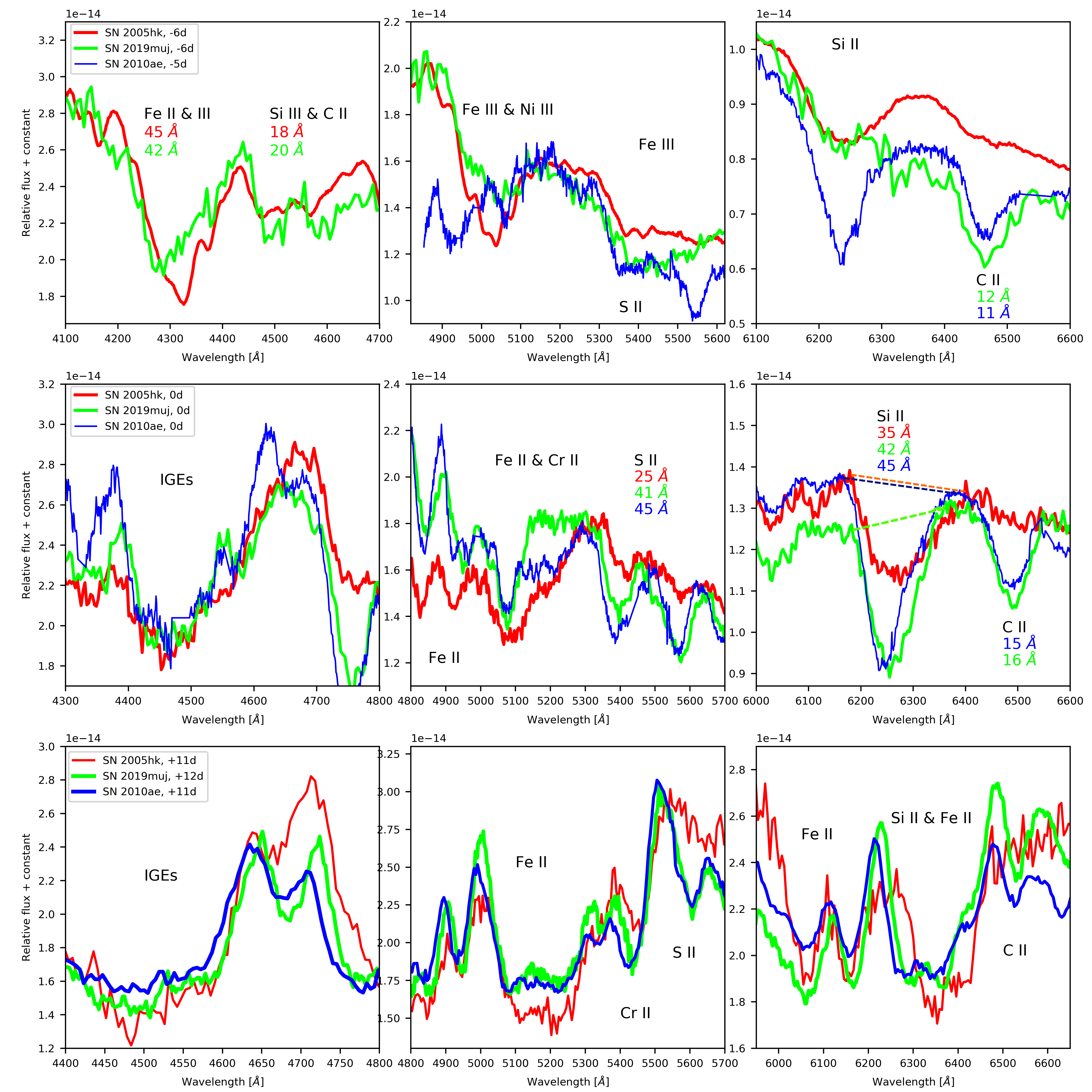}
    \caption{Zoomed insets of the spectra of SNe 2005hk, 2010ae, and 2019muj plotted in Fig. \ref{fig:comparison_iax}. For better comparison, the spectra of SNe 2005hk and 2010ha are scaled to those of SN 2019muj, and also shifted to match the photospheric velocity of SN 2019muj estimated from the Doppler-shift of the C II $\lambda$6580 absorption. The black labels show the ions forming the prominent absorption features. In some cases, where the emission peaks provide clear delimiters for the spectral feature, the pseudo-equivalent widths of the lines are also estimated and shown with the corresponding color codes. Examples for the different pseudo-continuum are indicated with dashed lines in the plot of the Si II $\lambda$6355 feature in the middle row.}
    \label{fig:comparison_lines}
\end{figure*}

\section{Discussion and results}
\label{discussion}

\subsection{Photometric analysis}
\label{photometry}

The ATLAS photometry provides a reliable and deep non-detection limit ($o>19.7$ mag, using 3$\sigma$ upper limit) on MJD $58\,696.58$ and a marginal (2$\sigma$ significance) detection on MJD $58\,700.52$ (as shown in Fig. \ref{fig:sn19muj_photometry_ATLAS} and Tab. \ref{tab:photometric_data_ATLAS}). Assuming the so-called expanding fireball model, the observed flux increases as a quadratic function of time ($F \sim t^{2}_\rmn{exp}$) before the epoch of maximum light \citep{Arnett82, Riess99, Nugent11}. Note that because of the lack of coverage in the first few days after explosion, the fitting of different LCs may deliver a different $t_\rmn{exp}$. To determine the time of the explosion more accurately, we simultaneously fit the {\textit g-}, {\textit c-} and {\textit o}-band LCs including the earliest observation and the best pre-maximum time resolution. The fitting function is:
\begin{equation}
F_\rmn{g} = a \cdot \left(t - t_{\rmn{exp}}\right)^{n},
\label{eq:fireball}
\end{equation}
where $a$ is a fitting parameter, while $n=2$ is kept fixed at first. The resulting date of explosion is found to be t$_\rmn{0}=58\,694.4$ MJD, which seems too early regarding to the constraints provided by the {\textit o}-band LC. However, another critical aspect of this analysis is the choice of power-law index. Detailed studies reported different power-law indices for samples of normal SNe Ia with mean values of n = 2.20 -- 2.44 \citep{Firth15,Ganeshalingam11}. However, our pre-maximum dataset for SN 2019muj is not sufficient to handle $n$ as a free parameter, which would lead to highly unlikely results ($n=0.7$ with constrained date of MJD $58\,700.5$).

As another approach, a fixed value for $n < 2$ can be chosen. In the case of SN 2015H, \cite{Magee16} presented the most detailed pre-maximum LC for a SN Iax and found $n = 1.3$ as the best-fit power-law index. If we assume similarity to SN 2015H, and adopt the same vaue of $n$ for Eq. \ref{eq:fireball} constraining the date of explosion as MJD $58\,698.1$, which is a more realistic date and compatible with the ATLAS forced photometry. Note that using $n = 1.3$ is still an arbitrary choice and no meaningful uncertainty can be estimated. Thus, we do not claim MJD $58\,698.1$ as the explosion date of SN 2019muj, instead, we use it only for comparison with the results of other methods hereafter (see in Sec. \ref{spectroscopy} and \ref{bolometric}).

We fit the LCs with low-order polynomial functions from the first observation until two weeks after their maxima. The resulting LC parameters can be found in Table \ref{tab:photometric_fitting}. According to the peak absolute magnitude in V-band, SN 2019muj is less luminous than the majority of studied SNe Iax having peak M$_V$ greater than $-$17 mag. Between these more luminous members of the class and the few extremely faint objects (M$_V > -15$ mag), there is a gap of well-studied SNe Iax. According to the peak absolute magnitude ($M_\rmn{V} = -16.42$ mag), the closest relatives of SN 2019muj are SNe 2004cs and 2009J \citep{Foley13}. However, SN 2009J has only post-maximum spectral epochs, which are insufficient to characterize the chemical properties via abundance tomography. SN 2004cs is also categorized as an SN Iax based on its only spectrum obtained at day +42 day, and \cite{White15} argue that the object does not even belong to the Iax class. Thus, SN 2019muj provides a unique opportunity to link the moderately bright SNe Iax to the brighter ones.

 The estimated value of the decline rate is $\Delta m_{15} = 2.4$ mag in $B$-band and 1.0 mag in $r$-band. These values indicate a faster dimming compared to SN 2005hk \citep[1.56 mag in $B$-band, 0.7 mag in $r$-band][]{Phillips07} and all the more luminous SNe Iax with detailed photometric analysis \citep{Foley13}. At the same time, the decline rates of SN 2019muj are comparable with those of SNe 2010ae, 2008ha \citep[2.43 and 2.03 mag in $B$-band; 1.01 and 1.11 mag in $r$-band, respectively,][]{Stritzinger14} and 2019gsc \citep[1.14 in $r$-band][]{Tomasella20}, the less energetic explosions in the class. As a conclusion, the decline rate is not well correlated to luminosity over the full range of SNe Iax. 

The $B-V$ intrinsic colour evolution of SN 2019muj is compared again to a more luminous \citep[SN 2005hk,][]{Phillips07} and a fainter \citep[SN 2008ha,][]{Stritzinger14} object. All SNe Iax show similar colour evolution in general \citep{Foley13}: the originally blue colour curve gets redder with time until 10--15 days after $B$-maximum, then a nearly constant color phase starts. As can be seen in Fig. \ref{fig:color_comparison}, the $B-V$ evolution of SN 2019muj is almost identical to that of SN 2005hk over the first $\sim$60 days. The only minor difference is the slightly bluer colour of SN 2019muj during the constant phase starting at 15 days after $B$-maximum. However, the actual extent of the discrepancy is uncertain because of the highly scattered photometry at this phase. On the other hand, SN 2008ha is redder with $\sim$0.6--0.7 mag at every epoch before the peak. It shows a similar $B-V$ value of $\sim$1.25 mag as SN 2019muj during the constant phase, though we note the lack of observations for SN 2008ha between 25 and 40 days after maximum.

SN 2019muj is only the fourth SN Iax in the literature with UV-photometry obtained by Swift. In Fig. \ref{fig:swift_comparison}, the UVW2, UVM2 and UVW1 LCs are compared to those of SNe 2011ay, 2012Z and 2005hk. All three SNe are more luminous than SN 2019muj. As one can observe, the post-maximum LCs of SN 2019muj show a faster fading, which could be a sign of a steep density profile of the outer ejecta.

\begin{figure*}
	\includegraphics[width=15cm]{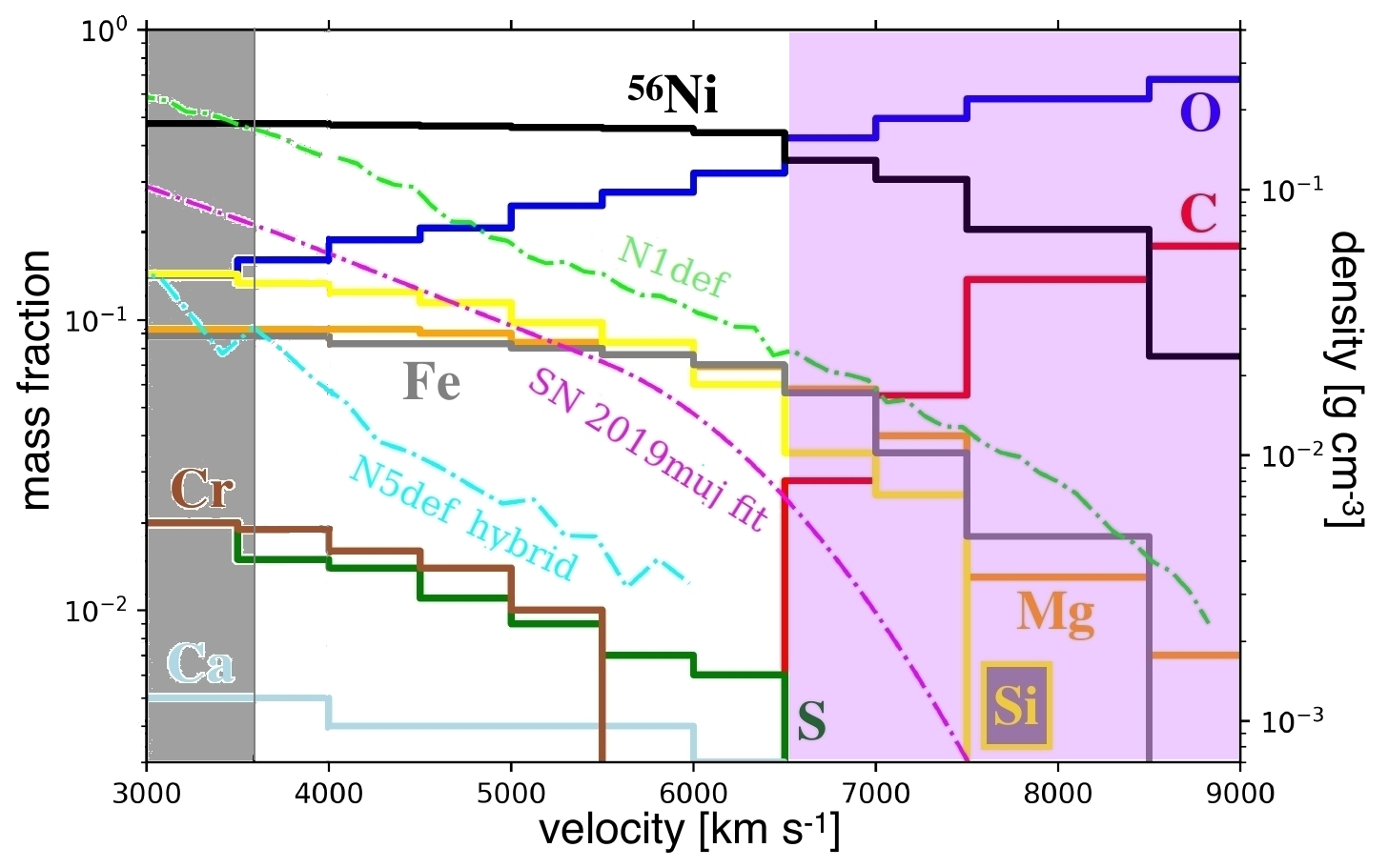}
    \caption{The adopted abundance template from \citet{Barna18} shifted with 6500 km s$^{-1}$ (i.e. the transition velocity between the $^{56}$Ni dominated inner and the O dominated outer regions). The green, cyan and magenta dashed lines represent the density profiles of the N1def \citep{Fink14}, the N5def\_hybrid \citep{Kromer15}, and the feasible fit {\tt TARDIS} model (this paper) scaled to t$_\rmn{exp}$ = 100 s, respectively. Note that the lack of earlier spectra and the steep cut-off in the adopted density profile above $\sim$6500 km s$^{-1}$ makes the constraints on abundance structure uncertain in this outer region (highlighted with purple area). Moreover, the latest TARDIS model has photospheric velocity of 3600 km s$^{-1}$, thus, we cannot test the chemical abundances below this limit (highlighted with grey area).}
    \label{fig:abundance_template}
\end{figure*}

\begin{figure*}
	\includegraphics[width=15cm]{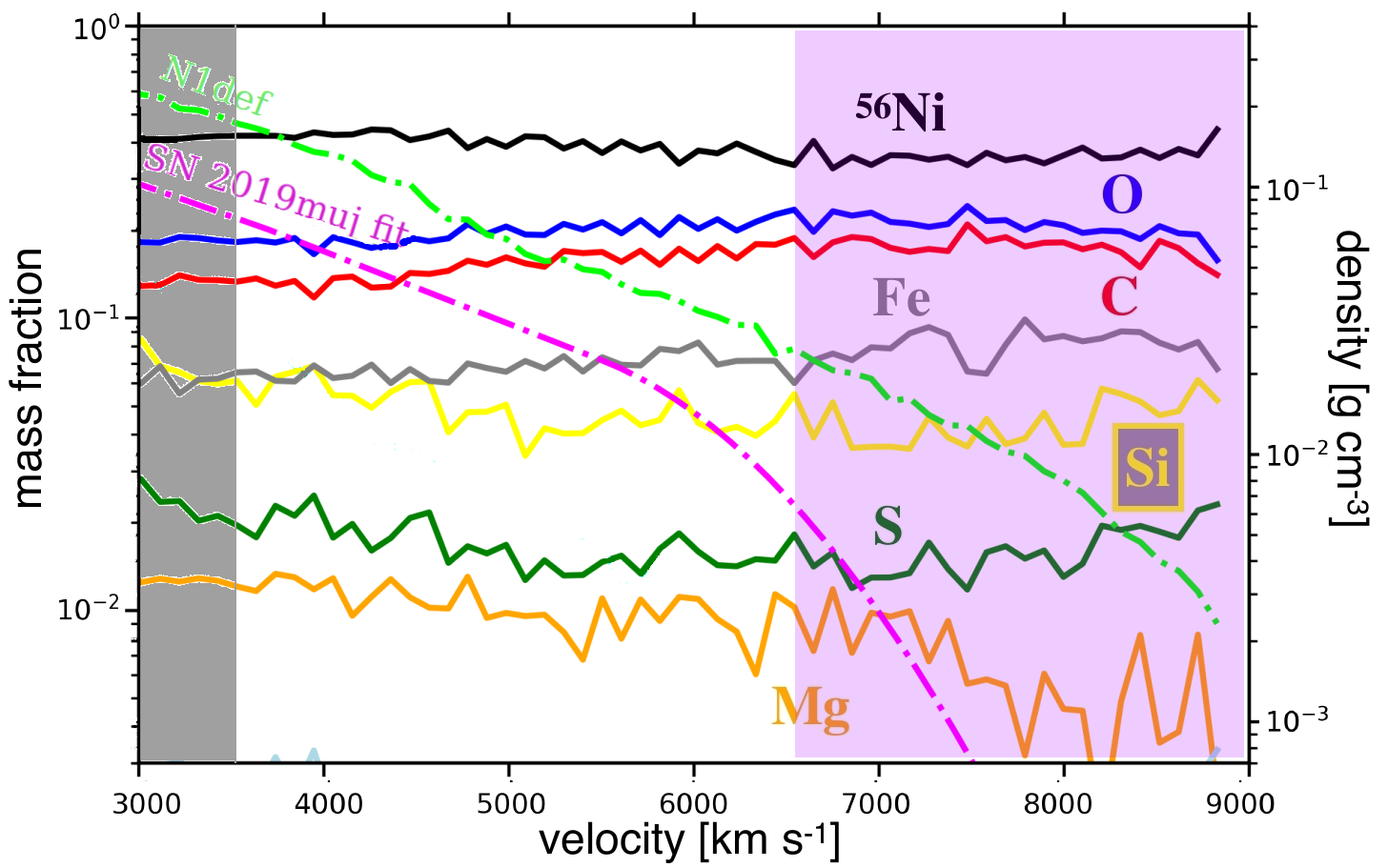}
    \caption{The chemical abundances of the N1def pure deflagration model \citep{Fink14} as comparison to Fig. \ref{fig:abundance_template}. The density profiles of N1def and our TARDIS model are also shown with dashed lines. Similarly to Fig. \ref{fig:abundance_template}, the uncertain (above $\sim$6500 km s$^{-1}$) and unseen (below 3600 km s$^{-1}$) regions are highlighted with purple and grey colors, respectively.}
    \label{fig:abundance_n1def}
\end{figure*}

\subsection{Spectroscopic analysis}
\label{spectroscopy}

Observed spectra of SN 2019muj are scaled and color-matched to the observed broad-band photometry. The resulting series of spectra are plotted in Fig. \ref{fig:sn19muj_spectroscopy}, while the log of each observation is listed in Tab. \ref{tab:spectroscopic_data}. 

In Fig. \ref{fig:comparison_iax}, the spectra of three objects are compared to those of other SNe Iax that have spectra close in time to SN 2019muj and that provide examples of more and less luminous events with respect to SN 2019muj. Although the same spectral lines appear in all objects, their relative optical depths as well as the shape of the continuum vary from object to object. For better comparison, some spectral features at each epochs are displayed in the insets of Fig. \ref{fig:comparison_lines}.

At 6 days before the maximum, SN 2019muj shows only weak signs of IMEs. The spectrum over 5500 \r{A} is nearly featureless, with C II $\lambda6580$ as the only easily identifiable line. The shorter wavelengths are dominated by wide Fe \begin{small}II\end{small} and Fe \begin{small}III\end{small} features, which, together with the blue continuum, suggest the presence of a hot medium ($T >12\,000$ K) close to the assumed photosphere. The appearance of the prominent IGE lines (e.g features at $\sim$4300 and $\sim$5000 \r{A}), as well as their pseudo-equivalent widths (pEW) are strikingly similar to those of SN 2005hk \citep{Phillips07} at this epoch. Note that the pEW-measurements \citep[for a brief description see e.g.][]{Childress13} adopted mainly for normal SNe Ia are questionable in the case of SNe Iax, because of the the severe overlaps of the wide absorption features.

At the same time, SN 2010ae \citep{Stritzinger14}, the extremely faint SN Iax involved in this comparison, shows strong Si \begin{small}II\end{small} $\lambda6355$ line and ``W'' feature of S \begin{small}II\end{small}, but restrained IGE absorption lines. These features, which typically require lower photospheric temperatures ($T < 11\,000$ K), are not identifiable in the spectrum of SN 2019muj. All these attributes make probable that SN 2019muj tends to be more similar to the more luminous SNe Iax compared to the subluminous objects of the Iax class at the pre-maximum epochs. The only common feature in both spectra is the significant C II $\lambda6580$ line, which indicates relatively high carbon abundance in the outermost region of SN 2019muj. However, the limited wavelength range of the spectrum of SN 2010ae (and the lack of spectra of other extremely faint SNe Iax at this epoch) makes this comparison incomplete.


The match changes around the maximum when the structures and the estimated pEWs of the IME features of SN 2019muj resemble more closely those of SN 2010ae, than those of SN 2015hk. The strength of the C II $\lambda6580$ feature still provide a good match with that of SN 2010ae, while this feature is barely identifiable in the spectrum of SN 2005hk. At the same time, the IGE dominated absorption feattures below 6000 \r{A} show a diverse nature. The pseudo-emission peaks between 4300 and 4700 \r{A} are in better agreement with the same features of SN 2005hk, while the more narrow Fe II absorption lines between 4700 and 5150 \r{A} are more similar to those of SN 2010ha. At $\sim$12 days after the maximum, the whole range spectral range of SN 2019muj is nearly a perfect match to that of SN 2010ae, while the SN 2005hk spectrum shows several more dominant pseudo-emission peaks (e.g. at 4700 and 5700 \r{A}.)

The different nature of SN 2019muj before and after maximum light might be the result of the steeper density structure. The thin outer region allows insight to the deeper, thus, hotter layers, which changes as the photosphere starts forming in the more dense regions. Such density structure may resemble more with that of the more luminous objects in the outer region with a cut-off at the highest velocities \citep[see the N5def and N20def deflagration models and the abundance tomographies of SNe Iax,][]{Fink14,Barna18}, and the profile quickly increase toward the innermost regions \citep[see the N1def and the hybrid N5def deflagration models,][]{Fink14,Kromer15}. It can not be stated whether this behavior is regular among the moderately luminous SNe Iax, or not, without studying more objects with approximately the same luminosity. The best candidates for this purpose are SNe 2014ck \citep{Tomasella16} and 2015H \citep{Magee16} so far, both SNe had a relatively moderate luminosity (still a magnitude brighter than SN 2019muj) and detailed observational data. However, constraining the density profile (see below) of multiple objects is beyond the scope of this study.

\begin{figure*}
	\includegraphics[width=18cm]{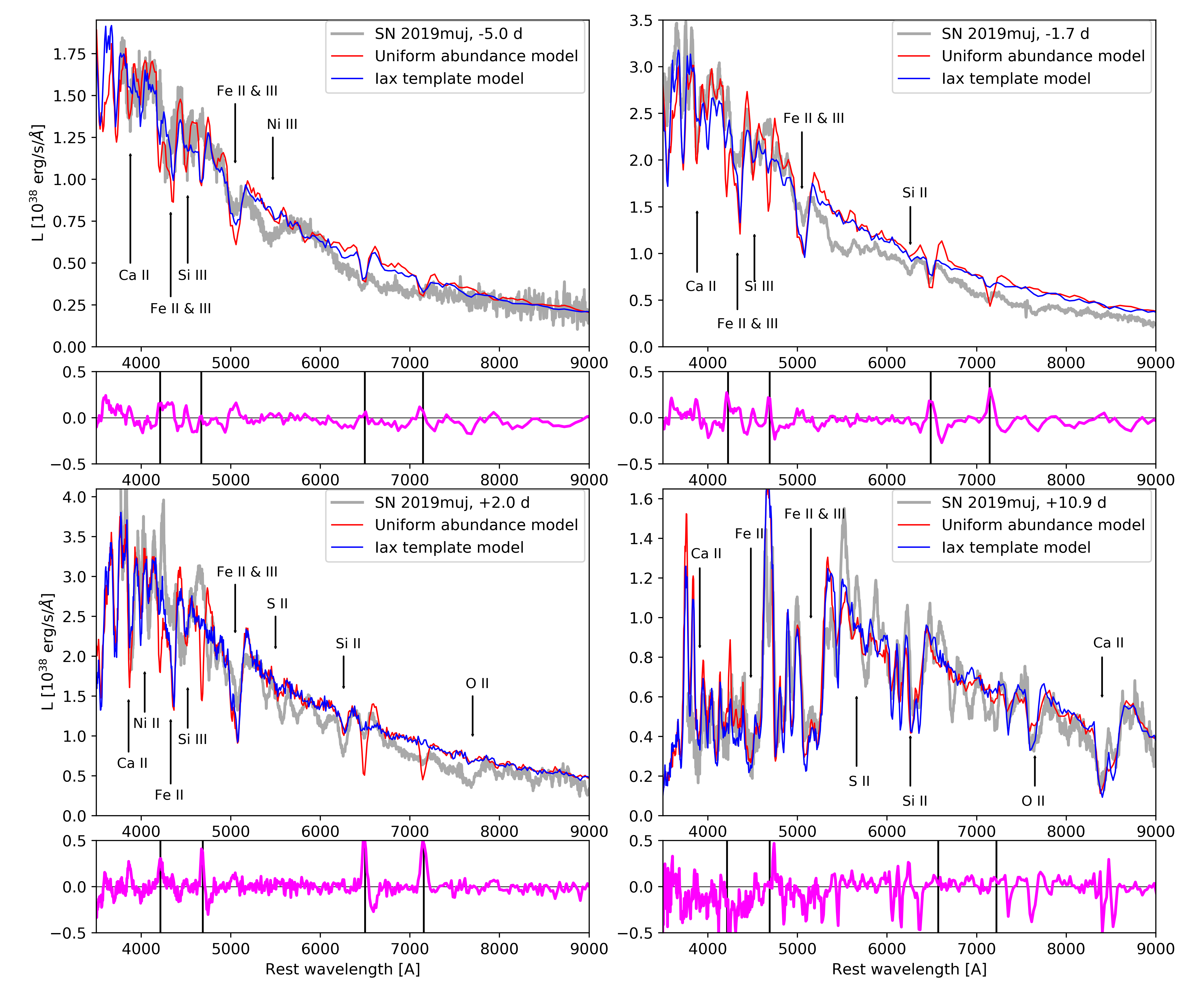}
    \caption{Four of the observed spectra of SN 2019muj obtained around maximum (grey), corrected for reddening and redshift, the {\tt TARDIS} synthetic spectra assuming the abundance template from \citet{Barna18} (blue) and the impact of assuming uniform abundances based on the N1def model from \citet{Fink14} (red). The purple residuals show the difference between the two synthetic spectra in each panel. Vertical lines indicate the absorption minima of the C II lines $\lambda4268$, $\lambda4746$, $\lambda6580$ and $\lambda7234$ in the residual plots. Key spectral lines formed by other ions in the template model are indicated by black arrows. }
    \label{fig:sn19muj_tardis}
\end{figure*}

\begin{figure*}
	\includegraphics[width=17.0cm]{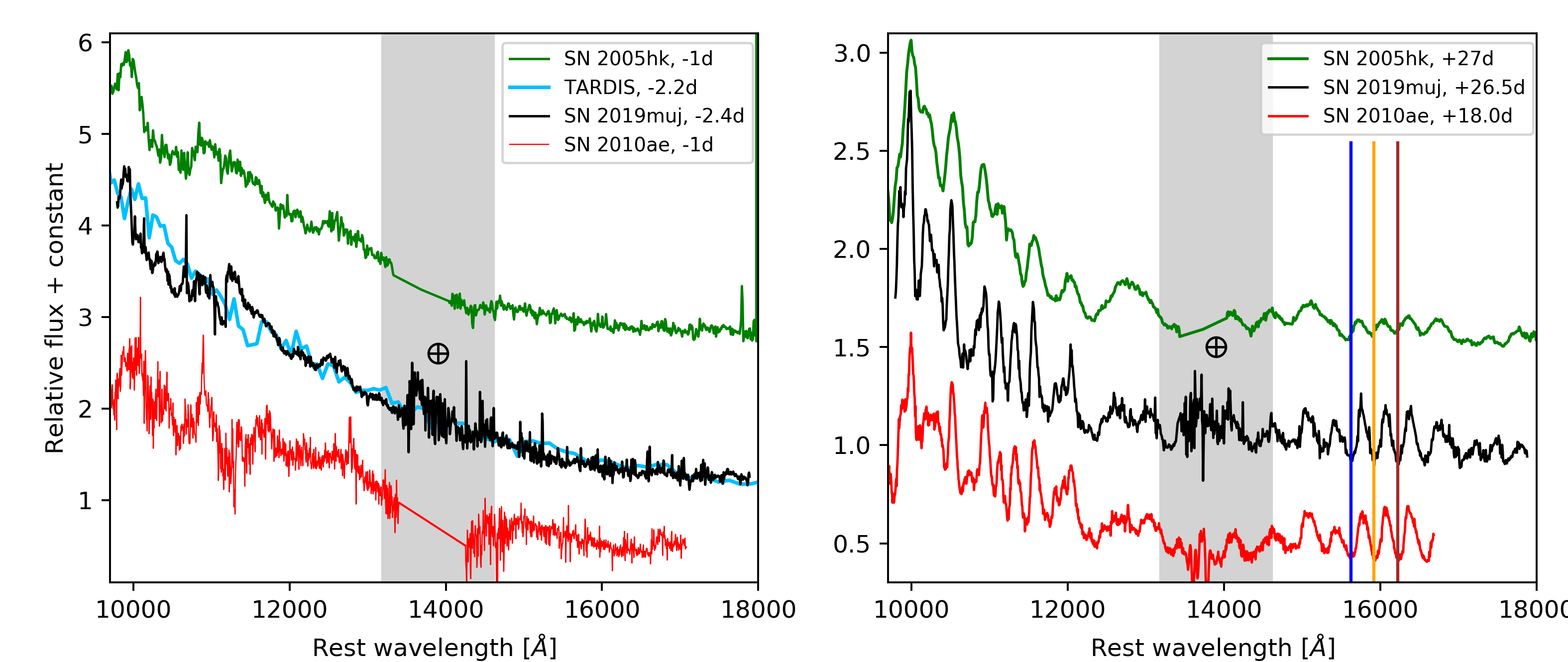}
    \caption{The two infrared spectra obtained with Gemini-South/FLAMINGOS-2 spectrograph, compared to to Type Iax SNe 2005hk \citep{Kromer13} and 2010ae \citep{Stritzinger14} obtained at similar epochs. The TARDIS synthetic spectrum fitted to the optical spectrum and extended to the longer wavelengths is also plotted in the first panel. The epochs show the days with respect to $B$-band maximum. The blue, orange and brown vertical lines show the absorption minima of the Co II $\lambda15795$, $\lambda16064$ and $\lambda16361$ lines for the second epoch. The wavelength range heavily affected by major telluric absorption lines are marked with grey stripe.}
    \label{fig:ir_spectra}
\end{figure*}

In order to test the physical structure of the ejecta, we fit the spectral time series with synthetic spectra calculated via the 1D Monte Carlo radiative transfer code \begin{small}TARDIS\end{small} \citep{Kerzendorf14, Kerzendorf19, Vogl19} in multiple ways. First, we use the abundance template proposed by \cite{Barna18} for the sample of more luminous SNe Iax. The template (see Fig. \ref{fig:abundance_template}) is paramterized by a velocity shift of the abundance structure, measured by the velocity where the $^{56}$Ni abundance becomes dominant. In the case of 2019muj, we use this velocity shift of the abundance template as a free parameter for fitting the spectra. The assumed density function is the same as those used by \citet[note that the related Eq.~1 in the cited paper is incorrect]{Barna18} for reproducing the density structure of various deflagration models of \cite{Fink14}. The density function contains an exponential inner part, which continues with a shallow cut-off past the velocity $v_\rmn{cut}$:
\begin{equation}
\rho (v,t_{\rmn{exp}}) = 
\begin{cases}
\rho_0 \cdot \left(\frac{t_{\rmn{exp}}}{t_0}\right)^{-3} \cdot \exp\left({-\frac{v}{v_0}}\right)
& \text{for } v \leq v_\rmn{cut} \\
\rho_0 \cdot \left(\frac{t_{\rmn{exp}}}{t_0}\right)^{-3} \cdot \exp\left({-\frac{v}{v_0}}\right) \cdot
8^{-\frac{8(v - v_\rmn{cut})^2}{v^{2}_\rmn{cut}}}
& \text{for } v > v_\rmn{cut}
\end{cases}
\end{equation}  
where $\rho_\rmn{0}$ is the reference density at the reference time of $t_\rmn{0} = 100$ s, $v$ is the velocity coordinate and $v_\rmn{0}$ is reference velocity, which defines the slope of the density profile. In our fitting strategy, we use $\rho_\rmn{0}$, $v_\rmn{0}$ and $v_\rmn{cut}$ as free parameters, while the date of the explosion is adopted from the LC analysis. Luminosities and photospheric velocities are constrained individually for each spectrum. 

Note that our approach of spectral fitting is only a simplified version of the technique called abundance tomography \citep{Stehle05}, because the chemical abundances are controlled via the template. Moreover, our fitting strategy does not result in a `best-fit' model, because the whole parameter-space cannot be fully explored manually. Instead, we aim to find a feasible solution for the main characteristics of SN 2019muj assuming physical and chemical structure which reproduced the main spectral features of several epoch in the case of more luminous SNe Iax.

We find a feasible solution assuming explosion date of MJD $58\,697.5 \pm 0.5$, which is in good agreement with the previously estimated value from the early LC fitting (MJD $58\,698.1$). The core density is assumed to be $\rho_\rmn{0}$ = 0.55 g cm$^{-3}$, while the slope of the function in the inner region is $v_\rmn{0}$ = 1800 km s$^{-1}$. The density profile starts to deviate from the pure exponential function above $v_\rmn{cut}$ = 5500 km s$^{-1}$. In accord with the expectations based on the loose correlation between the ejecta structure and the luminosity, these parameters are lower for SN 2019muj compared to those of previously analyzed SNe Iax \citep{Barna18}. The constructed density profile also falls short of reaching the density function of the N1def model \citep{Fink14}, but stays above the density function of the N5def\_hybrid model \citep{Kromer15}, directly developed for a possible explanation of the extremely faint SNe Iax (Fig.~\ref{fig:abundance_template}). Note that the photospheric velocity of the latest TARDIS model of this analysis limits the part of the ejecta we can study. Thus, we cannot gain any knowledge neither about the density or the chemical abundances below 3600 km s$^{-1}$.

As it was indicated in Sec. \ref{sec:introduction}, the impact of the outer regions on the synthetic spectrum strongly depend on the choice of density at high velocities. The adopted density profile has a steep cut-off above $\sim$6500 km s$^{-1}$ (see in Fig. \ref{fig:abundance_template}), which makes the abundance structure highly uncertain in this region. Thus, one should test, whether the abundances of specific chemical elements are indeed sensitive to the outermost velocity region. After constraining the set of fitting parameters for the abundance template, we calculate synthetic spectra with constant abundances using the same physical properties. The mass fractions of these models are adopted from the N1def model of \cite{Fink14} (see in Fig. \ref{fig:abundance_n1def}), averaging and normalizing the abundances of the chemical elements listed in Fig. \ref{fig:abundance_template}. The comparison between the two sets of synthetic spectra is plotted in Fig. \ref{fig:sn19muj_tardis}. 

The template models show a relatively good agreement with the observed spectra, especially a week after maximum light, where the synthetic spectra reproduce almost all the absorption features. At the earliest epochs, an excess of carbon is found despite the mass fraction decrease inward of the photosphere. This suggests that stronger limitation of carbon abundance is required to produce the spectral lines correctly. Around maximum light, the absorption features of IMEs do not or just slightly appear, which may require a more complex treatment of the temperature and density profile.

At the same time, the spectra calculated from the constant abundance model match similarly well to the observed data, apart from the spectral features of C II (see Fig. \ref{fig:sn19muj_tardis}). Due to the steep cut-off in the adopted density profile, testing the ejecta structure far over the photosphere of the first epoch ($v_\rmn{phot}=6\,000$ km s$^{-1}$) is highly uncertain as it is indicated in Figs. \ref{fig:abundance_template} and \ref{fig:abundance_n1def}. In the regions probed in our work we cannot tell the difference between the stratified and constant abundance model as our spectra do not go early enough, except in the case of carbon. As it is shown in Fig. \ref{fig:sn19muj_tardis}, carbon forms too strong absorptions in the case of the constant abundance model around the maximum light. This exceed in the uniform abundance models would result from the relatively high $X(C) \approx 0.13$ mass fraction below 6000 km s$^{-1}$ in the constant profile. This confirms the assumption of \cite{Barna18} that a significant amount of unburned material in the ejecta of SNe Iax is likely found only in the outermost layers.


As a result, our analysis seems to favour the implied a stratified structure over the constant abundance model, as stratified abundance profile is required to reproduce the main spectral features of carbon. In the case of other chemical elements, the differences between the two models can be probed only with earlier spectral observations. Apart from carbon, the uniform abundance profiles of the N1def model describe well the chemical structure of SN 2019muj in general. Considering the possible explosion scenarios (see Sec. \ref{sec:introduction}), which can explain some of the observables of SNe Iax, the constrained abundance profiles of SN 2019muj may not contradict strongly with the predictions of the pure deflagration scenario, but definitely require some further tuning in the hydrodynamic models at least in the case of unburned material.

\begin{figure}
	\includegraphics[width=9.0cm]{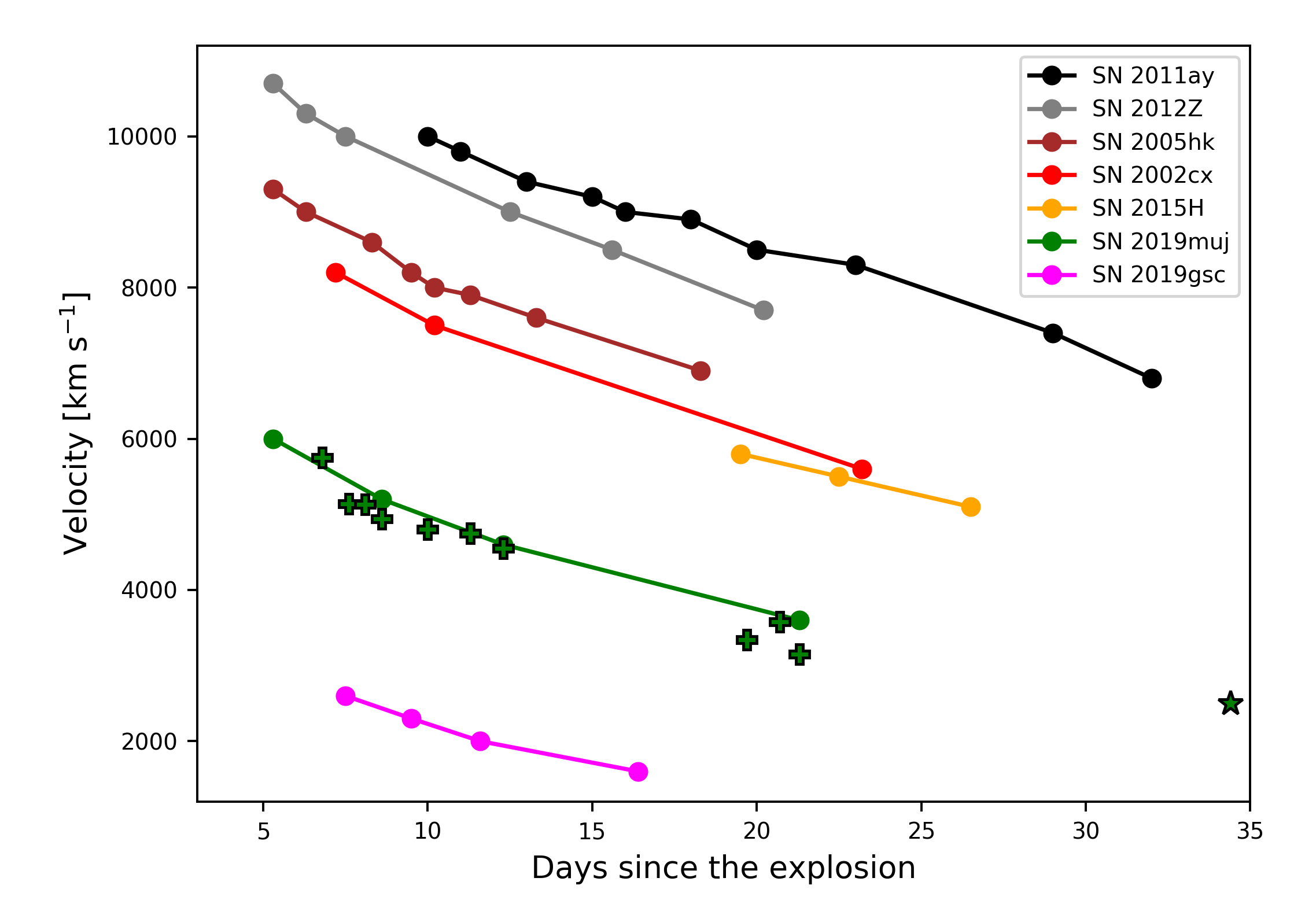}
    \caption{The evolution of photospheric velocity of SN 2019muj ($M_\rmn{V} = -16.42$ mag) and five relatively bright SNe Iax \citep{Barna18} with peak absolute brightness covering a range from $-$17.3 (SN 2015H) to $-$18.4 mag (SN 2011ay). The velocities (illustrated with filled circles) were constrained via abundance tomography analysis performed with {\tt TARDIS}. The photospheric velocities of the extremely faint SN 2019gsc ($M_\rmn{V} \sim -13.8$ mag) constrained via spectral fitting with {\tt TARDIS} are also plotted \citep{Srivastav20}. The crosses (between 5 and 221 days) and the star (at 34.4 days) show the photospheric velocity estimated from the shift of the absorption minima of Si II $\lambda$6355 and the NIR Co II lines, respectively.}
    \label{fig:sn19muj_vel}
\end{figure}

The self-consistent \begin{small}TARDIS\end{small} models also provide us an estimation of the photospheric velocity based on the simultaneous fits of several spectral lines and the flux continuum. The evolution of v$_\rmn{phot}$ is monotonically decreasing, with a steeper descent before the moment of the maximum light. The derived behaviour of v$_\rmn{phot}$ (see Fig. \ref{fig:sn19muj_vel}) shows good agreement with the average velocity estimated from the blueshift of the Si \begin{small}II\end{small} $\lambda$6355 absorption minima. Moreover, the v$_\rmn{phot}$ of the Co \begin{small}II\end{small} feature in the NIR regime at +26.5 days also supports this velocity evolution. The velocity profile of SN 2019muj fits into the trend of other SNe Iax, which showed a relation between the peak luminosity and the value of v$_\rmn{phot}$ at V-maximum \citep{McClelland10,Foley13}. This result suggests that this correlation is tight not only for the most luminous SNe Iax, but probably for the whole class. However, to achieve more general conclusions regarding to the luminosity-velocity relation, more SNe Iax \cite[including the reported outliers, e.g. SNe 2009ku and 2014ck,][]{Narayan11,Tomasella16} have to be studied with the same technique.

\begin{figure}
	\includegraphics[width=9.0cm]{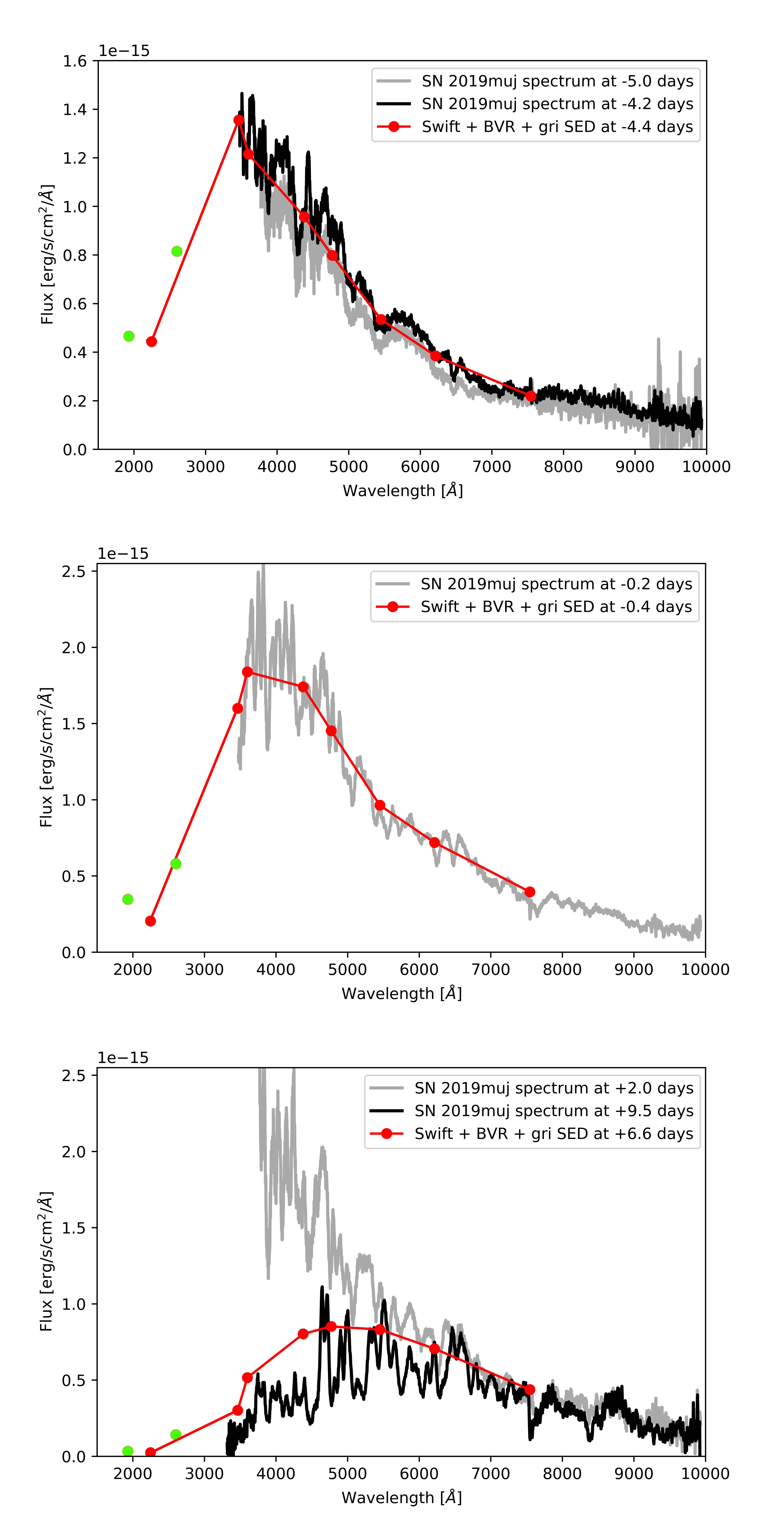}
    \caption{The SED of SN 2019muj calculated from UV/optical photometry at three different epochs, compared to the spectra obtained at the closest epochs (grey and black). The epochs indicate the time with respect to $B$-band maximum (MJD 58\,707.8). The flux values estimated from the Swift UVW2 and UVW1 bands are represented by two green points on each panel.}
    \label{fig:sed_evol}
\end{figure}

\begin{figure}
	\includegraphics[width=9.0cm]{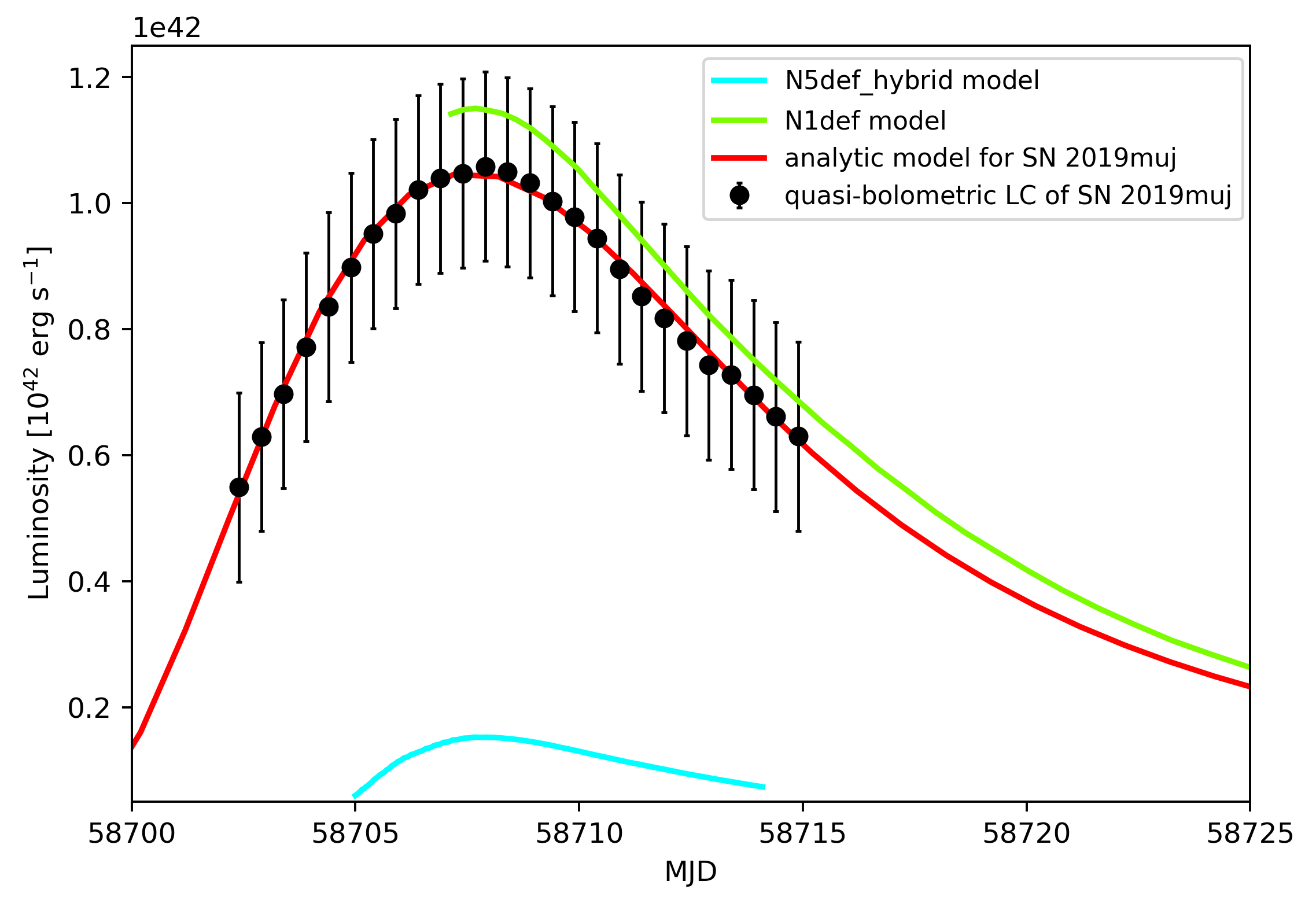}
    \caption{The bolometric LC of SN 2019muj and the best-fit Arnett-model described with Eq. \ref{eq:arnett}. The light curve peaks at MJD 58707.2. For comparison, the UVOIR LC of the N1def \citet[green,][]{Fink14} and N5def\_hybdrid \citet[cyan,][]{Kromer15} deflagration models are also plotted.}
    \label{fig:sn19muj_bol}
\end{figure}

The velocity estimation could be further improved by analyzing the near-infrared (NIR) spectral features. The two spectra of SN 2019muj obtained in the NIR regime are shown in Fig. \ref{fig:ir_spectra}. There are only a few SNe Iax with near-infrared (NIR) spectra: such observations have been presented in the case of SNe 2012Z \citep{Stritzinger15}, 2005hk \citep{Kromer13}, 2008ge \citep{Stritzinger14}, 2014ck \citep{Tomasella16}, 2015H \citep{Magee16} and 2010ae \citep{Stritzinger14}. Although TARDIS modeling is unreliable for fitting spectra at infrared wavelengths due to the black body approximation, the synthetic spectra could be used for identifying spectral features. Thus, we shift the flux of the TARDIS synthetic spectrum created for the -1.7 day epoch to match with the -2.4 day NIR spectrum. The shape of the continuum is of the synthetic and observed spectra are surprisingly similar, despite that real flux of the the synthetic spectrum is scaled with approximately an order of magnitude. However, the synthetic spectrum shows very weak spectral lines, which mostly overlap each other, making the precise line identification unfeasible. Note that the epoch of the latest optical spectrum fit with our TARDIS is 10.9 days after B-band maximum, thus, the comparison for the second NIR spectrum is not possible.

In Fig. \ref{fig:ir_spectra}, we also compare the two NIR spectra of SN 2019muj to those of SNe 2005hk and 2010ae obtained at similar epochs. Although, even the two comparative spectra are very similar at few days before the maximum light \citep[it was also shown by][]{Stritzinger14}, there are some differences between 10\,000 and 12\,000 \r{A} dominated by Mg II and Fe II lines \citep{Tomasella16}, where the absorption features are stronger, especially those at $\sim$11\,300 \r{A}, which barely noticeable in the spectrum of SN 2019muj. These properties make SN 2019muj resembling to the more luminous SN 2005hk at the earlier epoch. Similarly to the optical spectral evolution (Fig. \ref{fig:comparison_iax}), SN 2019muj at NIR wavelengths also turns to be more similar to the fainter SN 2010ae, as their spectra obtained at few weeks later (+26.5 and +18 days, respectively) are almost identical. At these epochs, the most prominent NIR features are the set of Co \begin{small}II\end{small} lines, which appear with significantly lower relative strength in the spectrum of SN 2005hk. The Co \begin{small}II\end{small} features do not overlap and allow us to measure the photospheric velocity with at post-maximum epochs \citep{Stritzinger14,Tomasella16}. Among the reported lines, Co \begin{small}II\end{small} $\lambda$15795, $\lambda$16064 and $\lambda$16361 can be easily identified in the later NIR spectrum of SN 2019muj. Based on the blueshifts of the absorption minima, the photospheric velocity is $v_\rmn{phot} = 2500$ km s$^{-1}$ at $+$26.5 days after $B$-band maximum light. We compare this value with other methods to investigate the evolution of the photosphere below.

\subsection{Bolometric light curve and SED}
\label{bolometric}

We calculated F$_\lambda$ flux densities from the {\it Swift}, $UBV$ and $gri$ dereddened magnitudes based on \cite{Bessel98} and \cite{Fukugita96}, respectively, and created SEDs for several epochs. Note that {\it Swift} wide band magnitudes are not used for this calculation, because of the potentially high uncertainties (see Sec. \ref{photometric_obs}). For a given date, if SN 2019muj was not observed in some of the filters, we use the corresponding value from the fitting of a low-order polynomial function to the observed magnitudes. The SEDs are directly compared to the spectra (corrected for reddening) in Fig. \ref{fig:sed_evol}. Integrating the SED functions to the total flux and scaling it according to the assumed distance, a quasi-bolometric LC is estimated (Fig. \ref{fig:sn19muj_bol}). The derived luminosities are fit with the analytic LC model introduced by \cite{Chatzopoulos12}, which is based on the radioactive decay diffusion model of \cite{Arnett82} and \cite{Valenti08}. The function has the following form:
\begin{equation}
\begin{split}
L(t) &= M_\rmn{Ni} \cdot \exp\left(-x^2\right) \cdot \left[1 - \exp\left(-A / t_\rmn{exp}^2\right) \right] \\
&  \cdot \Bigl[(\epsilon_\rmn{Ni} - \epsilon_\rmn{Co}) \int_{0}^{x} 2z \cdot \exp\left(z^2 - 2zy \right) dz \\
& + \epsilon_\rmn{Co} \int_{0}^{x} 2z \cdot \exp\left(z^2 - 2zy + 2zs\right) dz \Bigr] 
\end{split}
\label{eq:arnett}
\end{equation}
where $t_\rmn{exp}$ is the time since the explosion, M$_\rmn{Ni}$ is the radioactive nickel mass produced in the explosion, while $\epsilon_\rmn{Co}$ and $\epsilon_\rmn{Ni}$ are the energy generation rates by the decay of the radioactive isotopes of cobalt and nickel. Gamma-ray leakage from the ejecta is considered in the last term of the equation, where $A = (3\kappa_\gamma M_\rmn{ej}) \; / \;(4 \pi v^2)$ includes the gamma-ray opacity $\kappa_\gamma$, the expansion velocity $v$ and the mass of the ejecta $M_\rmn{ej}$. The additional parameters are $x = t \;/\; t_\rmn{d}$, $y = t_\rmn{d} \;/\; (2 t_\rmn{Ni})$, $z = 1\,/ \,2t_\rmn{d}$, and $s=t_\rmn{d} \cdot (t_\rmn{Co} - t_\rmn{Ni}) / (2 t_\rmn{Co} t_\rmn{Ni})$, where t$_\rmn{d}$ is the effective diffusion time.

The fit of the bolometric LC is plotted in Fig. \ref{fig:sn19muj_bol}, while the best-fit parameters are listed in Table \ref{tab:bolometric_fitting}. The date of explosion is found as MJD $58\,698.2 \pm 0.5$. This is in good agreement with the date estimated from the spectral fitting (MJD $58\,697.5 \pm 0.5$) and almost perfectly matches with the estimated value from the early LCs (MJD $58\,698.1$). Accepting MJD $58\,698.2$ as opposed to the one provided by the fireball model allows us to adjust the power-law fit of the early LCs, yielding $n \approx 1.26$. The rise times of SNe Iax are significantly shorter than those of normal SNe Ia, and this trend is confirmed in SN~2019muj, with $t_\rmn{rise} = 9.6 \pm 0.6$ days in the $B$-band (see Tab.~\ref{tab:photometric_fitting}). SN~2019muj also fits well to the trend of SNe Iax showing a loose correlation between their rise times and peak absolute magnitudes \citep[see e.g.][for V- and r- band comparisons]{Tomasella16,Magee16}.

Based on the bolometric LC fit, the amount of $^{56}$Ni produced during the explosion (M$_\rmn{Ni} = 0.031 \pm 0.005$ M$_\odot$) is low, even within the Iax class. It is approximately half of the estimated nickel mass of SN 2015H \citep[0.06 M$_\odot$,][]{Magee16}, $\sim 30\%$ of that of SN 2014ck \citep[0.1 M$_\odot$,][]{Tomasella16} and only $\sim 14\%$ of that of the brightest type Iax SN 2011ay \citep[0.225 M$_\odot$,][]{Szalai15}. At the same time, this nickel mass is an order of magnitude higher than those of the extremely faint SNe 2008ha and 2019gsc \citep[0.003 $M_\odot$,][]{Foley13,Srivastav20}, which confirms the placement of SN 2019muj in the gap of the luminosity distribution of SNe Iax. Nevertheless, the initial $^{56}$Ni mass of SN 2019muj matches reltively well with the value of M$_\rmn{Ni} = 0.0345$ M$_\odot$ of the N1def pure deflagration model calculated by \cite{Fink14} via hydrodynamical simulations. The similarity to N1def occurs not just on the peak brightness, but also on the post-maximum evolution of the LC (Fig. \ref{fig:sn19muj_bol}). This match further supports the idea that despite the possible discrepancies in the chemical abundances, SN 2019muj is the product of the pure deflagration scenario.

Assuming a correlation between $M_\rmn{ej}$ and $M_\rmn{Ni}$ as in the model grid of \cite{Fink14}, we may set the $M_\rmn{ej}=0.0843$ M$_\odot$ of the N1def model as an upper limit for SN 2019muj. Note however that the synthetic LCs of these pure deflagration models decline too rapidly compared to the SNe Iax LCs. The increased transparency of the ejecta is probably the simple consequence of the insufficient ejecta masses in the hydrodynamic simulations.
Another method was proposed by \cite{Foley09} taking advantage that both the rise time and the expansion velocity are proportional to the ejecta mass \citep{Arnett82}, which leads to:
\begin{equation}
\begin{split}
M_\rmn{ej} \propto v_\rmn{exp} \times t_\rmn{rise}^2 .
\end{split}
\label{eq:arnett2}
\end{equation}
Assuming that the mean opacity of the ejecta is uniform among thermonuclear SNe, this simplification allows to estimate the ejecta mass by comparing the bolometric rise time ($t_\rmn{rise}=9.0$ days, see in Fig. \ref{fig:sn19muj_bol}) and expansion velocity (here we adopt $v_\rmn{exp}=5000$ km s$^{-1}$) to those of normal SNe Ia. Following the formula presented by \cite{Ganeshalingam12}, we estimate $M_\rmn{ej}=0.17 $ M$_\odot$, while the kinetic energy of the ejecta is $E_\rmn{kin} = 4.4 \times 10^{49}$ erg. The discrepancy between ejecta masses calculated by this method and that of N1def deflagration model assumed as upper limit for SN 2019muj is similar to the mismatch estimated by \cite{Magee16}. While the ejecta mass of SN 2019muj is comparable to the values estimated for the less energetic Type Iax SN 2019gsc \cite[$M_\rmn{ej} \approx 0.13 - 0.20$ M$_\odot$;][respectively]{Tomasella20, Srivastav20}, the kinetic energy is significantly higher than that of SN 2019gsc \cite[$E_\rmn{kin} \approx 1.0 - 1.2 \times 10^{49}$ erg;][respectively]{Srivastav20,Tomasella20}.

\section{Conclusions}
\label{conclusions}
We present the observations of SN 2019muj, a SN Iax that exploded in the nearby galaxy VV 525. After discovery, the object was followed by both {\it Swift} and ground-based optical telescopes, providing LCs in the UV regime, as well as in $UBVR$ and $gri$ bands. The spectral series was delivered by a wide collaboration of observatories starting at 0.5 days after the discovery (-5 days with respect to $B$-band maximum) and covering a period of 60 days. The estimated peak luminosity places SN 2019muj between the relatively bright and the extremely faint objects in the diverse Iax class of thermonuclear explosions. Such a transitional SN Iax has not previously been a subject of detailed follow-up and analysis.

The rise times constrained from fitting a power-law function of $n=1.3$ \citep[similarly to the case of SN 2015H,][]{Magee16} fit well to the correlations with the peak absolute magnitudes \citep{Tomasella16, Magee16}. However, the estimated decline rates deviate from the previously reported loose correlation and show more similarity with those of extremely faint objects like SNe 2008ha and 2010ae. The intrinsic B-V colour evolution of SN 2019muj is almost identical to that of SN 2005hk. The blue continuum rapidly changes after B-maximum and shows a nearly constant B-V colour in the next weeks. Compared to SN 2008ha, the intrinsic colour of SN 2019muj is significantly bluer in the first three weeks after the explosion. 

The transitional nature of SN 2019muj can also be observed in the evolution of the optical spectra. According to the comparisons of the spectra obtained at the same epochs, SN 2019muj shows more similarity with the more luminous SNe 2005hk and 2011ay a few days before the date of B-maximum. At +12 days, both the lines of IMEs and IGEs match well with that of SN 2010ae.

Based on the comparisons of the observables, SN 2019muj shows a better match with the more luminous SNe Iax before the moment of maximum light, and with the extremely faint ones at the post-maximum epochs. This kind of change in the evolution of SN 2019muj might be the impact of its steep density profile. In order to examine, whether this behavior is regular among the moderately luminous SNe Iax, the study of more objects with approximately the same luminosity is required, which is out of the scope of this paper.

The spectra were also analyzed via a simplified version of abundance tomography, where a stratified abundance template \citep{Barna18} is fit by shifting it in the velocity space. Physical parameters, like luminosity and photospheric velocity, as well as the exponential function of the density profile, have also been free parameters in the fitting process. The density structure of the model shows a slope of $v_0 = 1800$ km s$^{-1}$ with a steep cut-off above 5500 km s$^{-1}$. The inferred feasible synthetic spectra were compared to spectra calculated with constant abundances. We found that the stratified abundance profile does not improve significantly the goodness of the fit in the case of most of the elements. However, without earlier spectra we are unable to accurately test abundances and densities in the outermost layers, where the deflagration and stratified models diverse. In the regions probed in this work the stratified and deflagration models are very similar, with the exception of carbon, which element is not allowed below 6500 km s$^{-1}$ in the stratified model. Our results demonstrate the need for carbon stratification, which is not inline with the state-of-the-art deflagration models.

We constructed a quasi-bolometric LC of SN 2019muj based on the UV and optical photometry. The estimated luminosities were fit with an analytic model, which allowed us to constrain the mass of radioactive nickel produced in the explosion as $M_\rmn{Ni}$ = 0.031 M$_\odot$. This value is slightly lower than the $M_\rmn{Ni}$ predicted by the N1def pure deflagration model reported by \cite{Fink14}. However, even weaker explosions were simulated assuming pure deflagration under special circumstances \citep{Kromer15}, thus, the bolometric LC of SN 2019muj does not contradict with deflagration scenario leaving behind a bound remnant. The ejecta mass was calculated assuming a constant opacity, as in the case of normal SNe Ia, with a value of $M\rmn{ej} = 0.17$ M$_\odot$. The analysis also delivered another constraint on the date of explosion as MJD $58\,698.2 \pm 0.5$. This shows a good agreement with the values obtained from early LC fitting (MJD $58\,698.1$) and abundance tomography (MJD $58\,697.5 \pm 0.5$).

SN 2019muj is not the first moderately luminous SN Iax, as SNe 2015H and 2014ck from the brighter side have also started to bridge the luminosity gap in the Iax class. However, the even lower peak luminosity, the detailed pre-maximum observations, the long-time follow-up, make SN 2019muj a good candidate to represent the intermediate population of the class.
Moreover, all the estimated physical and chemical properties suggest that SN 2019muj is a transitional object in the SN Iax class. Based on the preliminary comparisons with more luminous and extremely faint objects, it is very likely that certain physical properties have a continuous distribution among SNe Iax. This could be the source of several tight correlations, e.g. the previously proposed connection between the peak luminosity and expansion velocity. Nevertheless, this conclusion does not contradict with the idea that all SNe Iax share the similar explosion scenario, the pure deflagration of a WD leaving behind a bound remnant.

The verification of these assumptions require the re-investigation of the previously reported outliers \citep[e.g. SN 2014ck,][]{Tomasella16} to understand the origin of the discrepancies. Moreover, further SNe Iax with moderate peak luminosities have to be studied to prove that these objects indeed bridge the gap between the more luminous and the extremely faint SNe Iax.

\section*{Acknowledgements}

We are grateful for M. Stritzinger and S. Valenti for providing the NIR spectra of SN 2005hk. 
Based on observations collected at the European Organisation for Astronomical Research in the Southern Hemisphere, Chile as part of PESSTO, (the Public ESO Spectroscopic Survey for Transient Objects Survey) ESO program ID 1103.D-0328. 
This work makes use of observations from the LCO network.
This work has made use of data from the Asteroid Terrestrial-impact Last Alert System (ATLAS) project. ATLAS is primarily funded to search for near earth asteroids through NASA grants NN12AR55G, 80NSSC18K0284, and 80NSSC18K1575; by products of the NEO search include images and catalogs from the survey area. The ATLAS science products have been made possible through the contributions of the University of Hawaii Institute for Astronomy, the Queen's University Belfast, and the Space Telescope Science Institute. 

BB was funded by the Czech Science Foundation Grant no. 19-15480S and the institutional project RVO:67985815. This work is part of the project Transient Astrophysical Objects GINOP-2-3-2- 15-2016-00033 of the National Research, Development and Innovation Office (NKFIH), Hungary, funded by the European Union. Research on type-Iax supernovae at Rutgers University (SWJ, YC-N, LK) is supported by NSF award AST-1615455. L.G. was funded by the European Union's Horizon 2020 research and innovation programme under the Marie Sk\l{}odowska-Curie grant agreement No. 839090. DAH, JB, DH, and CP are supported by NSF AST-1911225 and NASA grant 80NSSC19K1639. Research by DJS is supported by NSF grants AST-1821967, 1821987, 1813708, 1813466, and 1908972, and by the Heising Simons Foundation under grant \#2020-1864. MN is supported by a Royal Astronomical Society Research Fellowship. TWC acknowledges the EU Funding under Marie Sk\l{}odowska-Curie grant agreement No 842471. KM acknowledges support from ERC Starting Grant grant no. 758638. STFC 2018 - 2020 : SJS and SS acknowledges funding from STFC Grant Ref: ST/P000312/1. This work has been partially supported by the Spanish grant PGC2018-095317-B-C21 within the European Funds for Regional Development (FEDER). TMB was funded by the CONICYT PFCHA / DOCTORADOBECAS CHILE/2017-72180113. MG is supported by the Polish NCN MAESTRO grant 2014/14/A/ST9/00121. JDL acknowledges support from STFCvia grant ST/P000495/1. M.R.S.\ is supported by the National Science Foundation Graduate Research Fellowship Program Under Grant No.\ 1842400. D.A.C. acknowledges support from the National Science Foundation Graduate Research Fellowship under Grant DGE1339067.

 The UCSC team is supported in part by NSF grant AST-1518052, NASA/Swift grant 80NSSC19K1386, the Gordon \& Betty Moore Foundation, the Heising-Simons Foundation, and by a fellowship from the David and Lucile Packard Foundation to R.J.F. 
 
 This work includes data obtained with the Swope Telescope at Las Campanas Observatory, Chile, as part of the Swope Time Domain Key Project (PI: Piro, Co-Is: Drout, Phillips, Holoien, French, Cowperthwaite, Burns, Madore, Foley, Kilpatrick, Rojas-Bravo, Dimitriadis, Hsiao). We wish to thank Swope Telescope observers Jorge Anais Vilchez, Abdo Campillay, Yilin Kong Riveros, Nahir Munoz Elgueta and Natalie Ulloa for collecting data presented in this paper. 
 
 We are grateful to the SALT Astronomers and support staff who obtained the SALT data presented here.
 
 Based on observations obtained at the international Gemini Observatory,  under program GS-2019B-Q-236, which is managed by the Association of Universities for Research in Astronomy (AURA) under a cooperative agreement with the National Science Foundation. On behalf of the Gemini Observatory partnership: the National Science Foundation (United States), National Research Council (Canada), Agencia Nacional de Investigaci\'{o}n y Desarrollo (Chile), Ministerio de Ciencia, Tecnolog\'{i}a e Innovaci\'{o}n (Argentina), Minist\'{e}rio da Ci\^{e}ncia, Tecnologia, Inova\c{c}\~{o}es e Comunica\c{c}\~{o}es (Brazil), and Korea Astronomy and Space Science Institute (Republic of Korea).
 
 Research at Lick Observatory is partially supported by a generous gift from Google.
 
This research made use of \textsc{Tardis}, a community-developed software
package for spectral synthesis in supernovae \citep{Kerzendorf14,Kerzendorf19}. The development of \textsc{Tardis} received support from the Google Summer of Code initiative and from ESA's Summer of Code in Space program. \textsc{Tardis} makes extensive use of Astropy and PyNE.

\section*{Data Availability}

The data underlying this article will be available on WISeREP, at
\url{https://wiserep.weizmann.ac.il/object/12935}.







\clearpage
\onecolumn
\appendix
\section{Observational data}

\begin{center}
\begin{longtable}{ccccc}
\caption{Log of ATLAS photometry obtained in \textit{cyan}- and \textit{orange}-filters. The fluxes are weighted means of the individual 30 second exposures on each night, and the errors are the standard deviations of those fluxes. All mags are in AB system. Before MJD $58\,700$, we adopt Flux [mag] = -2.5 log$_\rmn{10}$ (3 $\times$ Flux\_error) + 23.9 as 3$\sigma$ upper limit.} \label{tab:photometric_data_ATLAS} \\
\hline
\multicolumn{1}{c}{\textbf{MJD}} & \multicolumn{2}{c}{\textbf{cyan-filter}} & \multicolumn{2}{c}{\textbf{orange-filter}} \\

& Flux [$\mu$Jy] & Flux [mag] & Flux [$\mu$Jy] & Flux [mag] \\

\hline
58686.54 & - & - & 13.92 (101.61) & > 17.7 \\
58688.60 & - & - & 6.89 (19.47)	& > 19.5 \\
58690.59 & - & - & -0.54 (16.26) & > 19.7 \\
58692.56 & - & - & -1.51 (15.38) & > 19.7 \\
58694.51 & 1.68 (4.51)	& > 21.0 & - & - \\
58696.58 & - & - & 1.87 (19.79) & > 19.5 \\
58700.52 & - & - & 22.52 (12.56) & 20.52 (0.60) \\
58702.53 & 392.09 (6.36) & 17.42 (0.02) & - & - \\
58704.53 & - & - & 589.18 (14.71) & 16.97 (0.03) \\
58706.54 &  937.08 (7.96) & 16.47 (0.01) & - & - \\
58708.47 & - & - & 820.62 (38.08) & 16.61 (0.05) \\
58712.52 & - & - & 912.76 (72.30) & 16.50 (0.09) \\
\hline
\hline
\end{longtable}
\end{center}

\begin{center}
\begin{longtable}{ccccccc}
\caption{Log of Swift photometry. } \label{tab:photometric_data_Swift} \\
\hline
\multicolumn{1}{c}{\textbf{MJD}} & \multicolumn{1}{c}{\textbf{UVW2}} & \multicolumn{1}{c}{\textbf{UVM2}} & \multicolumn{1}{c}{\textbf{UVW1}} & \multicolumn{1}{c}{\textbf{U}} & \multicolumn{1}{c}{\textbf{B}} & \multicolumn{1}{c}{\textbf{V}} \\ \hline 

58702.85	 & 	17.78 (0.11)	 & 	17.48 (0.08)	 & 	16.77 (0.09)	 & 	16.21 (0.08)	 & 	17.26 (0.08)	 & 	17.19 (0.13) \\	
58703.85	 & 	17.49 (0.11)	 & 	17.58 (0.08)	 & 	16.63 (0.08)	 & 	15.89 (0.07)	 & 	16.92 (0.08)	 & 	16.86 (0.11) \\	
58704.72	 & 	17.62 (0.10)	 & 	17.71 (0.09)	 & 	16.70 (0.08)	 & 	15.76 (0.07)	 & 	16.75 (0.07)	 & 	16.78 (0.10) \\	
58705.62	 & 	17.92 (0.11)	 & 	18.04 (0.10)	 & 	16.79 (0.08)	 & 	15.87 (0.07)	 & 	16.70 (0.07)	 & 	16.53 (0.09) \\	
58706.83	 & 	17.89 (0.11)	 & 	18.13 (0.11)	 & 	17.00 (0.09)	 & 	15.77 (0.07)	 & 	16.51 (0.07)	 & 	16.49 (0.09) \\	
58710.69	 & 	18.79 (0.16)	 & 	19.45 (0.24)	 & 	17.80 (0.12)	 & 	16.51 (0.08)	 & 	16.67 (0.07)	 & 	16.27 (0.09) \\	
58712.94	 & 	19.77 (0.28)	 & 	20.20 (0.34)	 & 	18.36 (0.16)	 & 	17.17 (0.10)	 & 	16.98 (0.08)	 & 	16.58 (0.10) \\	
58717.80	 & 	-	 & 	-	 & 	-	 & 	18.73 (0.24)	 & 	18.07 (0.12)	 & 	17.04 (0.13) \\	
58720.12	 & 	-	 & 	-	 & 	-	 & 	-	 & 	18.54 (0.16)	 & 	17.10 (0.14) \\	
58724.58	 & 	-	 & 	-	 & 	-	 & 	-	 & 	18.85 (0.16)	 & 	17.50 (0.20) \\	
58726.10	 & 	-	 & 	-	 & 	-	 & 	-	 & 	18.69 (0.24)	 & 	17.77 (0.28) \\	
58729.72	 & 	-	 & 	-	 & 	-	 & 	-	 & 	19.19 (0.21)	 & 	17.72 (0.17) \\	
58730.31	 & 	-	 & 	-	 & 	-	 & 	19.12 (0.25)	 & 	19.18 (0.19)	 & 	17.97 (0.18) \\	
58735.66	 & 	-	 & 	-	 & 	-	 & 	-	 & 	19.34 (0.32)	 & 	17.99 (0.28) \\
\hline
\hline
\end{longtable}
\end{center}

\pagebreak

\begin{center}
\begin{longtable}[c]{ccccc}
\caption{Log of ground-based photometry obtained in \textit{UBV} filters. } \label{tab:photometric_data_UBV} \\
\hline
\multicolumn{1}{c}{\textbf{MJD}} & \multicolumn{1}{c}{\textbf{U}} & \multicolumn{1}{c}{\textbf{B}} & \multicolumn{1}{c}{\textbf{V}} & \multicolumn{1}{c}{\textbf{Telescope}}  \\ \hline 
\endfirsthead

\multicolumn{5}{c}%
{ \tablename\ \thetable{} -- continued from previous page} \\
\hline
\multicolumn{1}{c}{\textbf{MJD}} & \multicolumn{1}{c}{\textbf{U}} & \multicolumn{1}{c}{\textbf{B}} & \multicolumn{1}{c}{\textbf{V}} & \multicolumn{1}{c}{\textbf{Telescope}}  \\ \hline 
\endhead

\hline \multicolumn{5}{r}{{Continued on next page}} \\ \hline
\endfoot

\hline \hline
\endlastfoot

58702.70	&	16.56	(0.04)	&	17.32	(0.02)	&	17.42	(0.03)	&	LCO	\\
58702.70	&	16.51	(0.03)	&	17.34	(0.02)	&	17.40	(0.02)	&	LCO	\\
58703.40	&		-	&	17.17	(0.01)	&	17.08	(0.01)	&	Swope	\\
58703.40	&	16.45	(0.04)	&	17.11	(0.02)	&	17.16	(0.02)	&	LCO	\\
58703.40	&	16.37	(0.04)	&	17.12	(0.01)	&	17.13	(0.02)	&	LCO	\\
58703.80	&	16.19	(0.03)	&		-	&		-	&	LCO	\\
58704.38	&		-	&	16.85	(0.01)	&	16.76	(0.01)	&	Swope	\\
58704.40	&	16.08	(0.04)	&	16.79	(0.02)	&	16.84	(0.02)	&	LCO	\\
58704.40	&	16.12	(0.05)	&	16.79	(0.02)	&	16.82	(0.02)	&	LCO	\\
58704.50 	&		-	&		-	&	16.81	(0.01)	&	Thacher	\\
58705.10	&	16.02	(0.02)	&		-	&		-	&	LCO	\\
58705.10	&	16.01	(0.02)	&		-	&		-	&	LCO	\\
58705.50 	&		-	&		-	&	16.61 	(0.02)	&	Thacher	\\
58706.00	&	15.88	(0.03)	&	16.49	(0.02)	&	16.52	(0.02)	&	LCO	\\
58706.00	&	15.87	(0.02)	&	16.49	(0.02)	&	16.49	(0.02)	&	LCO	\\
58706.40	&	15.88	(0.04)	&	16.52	(0.01)	&	16.50	(0.01)	&	LCO	\\
58706.40	&		-	&	16.49	(0.01)	&	16.50	(0.02)	&	LCO	\\
58706.50 	&		-	&		-	&	16.49 	(0.03)	&	Thacher	\\
58707.30	&	15.90	(0.05)	&	16.46	(0.02)	&	16.41	(0.02)	&	LCO	\\
58707.30	&	15.90	(0.05)	&	16.48	(0.02)	&	16.43	(0.01)	&	LCO	\\
58707.70	&	15.93	(0.02)	&	16.44	(0.01)	&	16.38	(0.01)	&	LCO	\\
58707.70	&	15.93	(0.03)	&	16.44	(0.01)	&	16.40	(0.01)	&	LCO	\\
58708.20	&	16.07	(0.02)	&		-	&		-	&	LCO	\\
58709.10	&	16.01	(0.03)	&	16.44	(0.02)	&	16.34	(0.02)	&	LCO	\\
58709.10	&	16.06	(0.03)	&	16.45	(0.02)	&	16.36	(0.02)	&	LCO	\\
58710.30	&	16.35	(0.08)	&	16.63	(0.03)	&	16.37	(0.03)	&	LCO	\\
58710.30	&	16.17	(0.10)	&	16.60	(0.03)	&	16.37	(0.02)	&	LCO	\\
58710.40	&	16.18	(0.03)	&	16.50	(0.01)	&	16.35	(0.02)	&	LCO	\\
58710.40	&	16.27	(0.02)	&	16.51	(0.02)	&	16.36	(0.02)	&	LCO	\\
58712.41	&		-	&	16.96	(0.01)	&	16.42	(0.01)	&	Swope	\\
58713.00	&	16.99	(0.05)	&	16.95	(0.03)	&	16.53	(0.03)	&	LCO	\\
58713.00	&	17.03	(0.06)	&	16.97	(0.03)	&	16.49	(0.02)	&	LCO	\\
58714.10	&	17.20	(0.10)	&	17.31	(0.02)	&	16.73	(0.02)	&	LCO	\\
58714.10	&	17.33	(0.03)	&	17.34	(0.02)	&	16.73	(0.02)	&	LCO	\\
58714.10	&	17.28	(0.03)	&		-	&		-	&	LCO	\\
58715.29	&		-	&	17.58	(0.01)	&	16.71	(0.01)	&	Swope	\\
58715.30	&	17.59	(0.04)	&	17.53	(0.02)	&	16.70	(0.02)	&	LCO	\\
58715.30	&	17.53	(0.03)	&	17.55	(0.02)	&	16.71	(0.01)	&	LCO	\\
58717.30	&		-	&	17.96	(0.02)	&	16.90	(0.02)	&	Swope	\\
58717.40	&		-	&	17.97	(0.02)	&	16.81	(0.02)	&	LCO	\\
58717.40	&	18.46	(0.05)	&	17.93	(0.03)	&	16.91	(0.02)	&	LCO	\\
58717.51 	&		-	&		-	&	16.87 	(0.01)	&	Thacher	\\
58718.30	&	18.33	(0.04)	&	18.11	(0.02)	&	17.04	(0.02)	&	LCO	\\
58718.30	&	18.29	(0.03)	&	18.13	(0.02)	&	17.01	(0.02)	&	LCO	\\
58719.39	&		-	&	18.31	(0.01)	&	17.12	(0.01)	&	Swope	\\
58720.43	&		-	&	18.54	(0.05)	&	17.21	(0.02)	&	Swope	\\
58724.30	&	19.14	(0.04)	&	18.88	(0.02)	&	17.59	(0.02)	&	LCO	\\
58724.30	&	19.16	(0.04)	&	18.87	(0.02)	&	17.56	(0.01)	&	LCO	\\
58725.39	&		-	&	18.92	(0.01)	&	17.62	(0.01)	&	Swope	\\
58726.38	&		-	&	18.99	(0.02)	&	17.66	(0.01)	&	Swope	\\
58727.33	&		-	&	19.03	(0.01)	&	17.74	(0.01)	&	Swope	\\
58727.50 	&		-	&		-	&	17.72 	(0.01)	&	Thacher	\\
58728.10	&	19.69	(0.10)	&	19.06	(0.03)	&	17.85	(0.02)	&	LCO	\\
58728.10	&	19.48	(0.08)	&	19.16	(0.03)	&	17.83	(0.02)	&	LCO	\\
58729.33	&		-	&	19.16	(0.01)	&	17.82	(0.01)	&	Swope	\\
58732.40	&	19.58	(0.17)	&		-	&		-	&	LCO	\\
58732.40	&	19.61	(0.16)	&	19.30	(0.03)	&	17.98	(0.02)	&	LCO	\\
58732.50	&		-	&	19.33	(0.03)	&	17.96	(0.02)	&	LCO	\\
58736.20	&	20.45	(0.29)	&	19.55	(0.05)	&		-	&	LCO	\\
58736.20	&	20.47	(0.31)	&	19.51	(0.05)	&		-	&	LCO	\\
58736.34	&		-	&	19.38	(0.02)	&	18.05	(0.02)	&	Swope	\\
58736.60	&	19.89	(0.23)	&	19.35	(0.06)	&	18.16	(0.03)	&	LCO	\\
58736.60	&	19.48	(0.23)	&	19.38	(0.05)	&	18.20	(0.03)	&	LCO	\\
58737.40	&		-	&	19.42	(0.02)	&	18.14	(0.01)	&	Swope	\\
58740.20 	&		-	&	19.24	(0.20)	&	18.11	(0.12)	&	LCO	\\
58740.20 	&		-	&	19.76 	(0.34)	&	17.98	(0.10)	&	LCO	\\
58744.30 	&		-	&	19.66	(0.10)	&	18.23	(0.04)	&	LCO	\\
58744.30 	&		-	&	19.49 	(0.08)	&		-	&	LCO	\\		
58745.33 	&		-	&	19.56 	(0.04)	&	18.42	(0.03)	&	Swope	\\
58746.34 	&		-	&	19.62 	(0.02)	&	18.31	(0.02)	&	Swope	\\
58748.40 	&		-	&	19.69	(0.05)	&	18.46	(0.02)	&	Swope	\\
58752.10 	&		-	&	19.67 	(0.03)	&	18.65	(0.03)	&	LCO	\\
58752.10 	&		-	&	19.70	(0.03)	&	18.60	(0.03)	&	LCO	\\
58752.33 	&		-	&	19.74	(0.01)	&	18.49	(0.01)	&	Swope	\\
58753.15 	&		-	&	19.78	(0.02)	&	18.62	(0.02)	&	Swope	\\
58754.27 	&		-	&	19.80	(0.02)	&	18.59	(0.01)	&	Swope	\\
58755.27 	&		-	&	19.73	(0.01)	&	18.58	(0.01)	&	Swope	\\
58758.40 	&		-	&	19.80	(0.05)	&	18.74	(0.03)	&	LCO	\\
58758.40 	&		-	&	19.86	(0.05)	&	18.64	(0.03)	&	LCO	\\
58759.28	&		-	&	19.80	(0.02)	&	18.66	(0.02)	&	Swope	\\

\end{longtable}
\end{center}

\begin{center}
\begin{longtable}{ccccccc}
\caption{Log of ground-based photometry obtained in \textit{ugriz} filters.} \label{tab:photometric_data_ugriz} \\
\hline

\multicolumn{1}{c}{\textbf{MJD}} & \multicolumn{1}{c}{\textbf{u}} & \multicolumn{1}{c}{\textbf{g}} & \multicolumn{1}{c}{\textbf{r}} & \multicolumn{1}{c}{\textbf{i}} & \multicolumn{1}{c}{\textbf{z}} & \multicolumn{1}{c}{\textbf{Telescope}}  \\ \hline 
\endfirsthead

\multicolumn{7}{c}%
{ \tablename\ \thetable{} -- continued from previous page} \\
\hline
\multicolumn{1}{c}{\textbf{MJD}} & \multicolumn{1}{c}{\textbf{u}} & \multicolumn{1}{c}{\textbf{g}} & \multicolumn{1}{c}{\textbf{r}} & \multicolumn{1}{c}{\textbf{i}} & \multicolumn{1}{c}{\textbf{z}} & \multicolumn{1}{c}{\textbf{Telescope}}  \\ \hline 
\endhead

\hline \multicolumn{7}{r}{{Continued on next page}} \\ \hline
\endfoot

\hline \hline
\endlastfoot

58702.40	&	-		&	17.40 (0.00)	&	-		&	-		&	-		&	Cassius	\\
58702.70	&	-		&	17.21 (0.01)	&	17.47 (0.02)	&	17.71 (0.05)	&	-		&	LCO	\\
58702.70	&	-		&	17.20 (0.01)	&	17.47 (0.02)	&	17.75 (0.05)	&	-		&	LCO	\\
58703.39	&	16.88 (0.01)	&	16.96 (0.01)	&	17.13 (0.01)	&	17.42 (0.01)	&	-		&	Swope	\\
58703.40	&	-		&	-		&	17.23 (0.01)	&	17.45 (0.02)	&	-		&	LCO	\\
58703.40	&	-		&	-		&	-		&	17.45 (0.02)	&	-		&	LCO	\\
58704.37	&	16.70 (0.01)	&	16.69 (0.01)	&	16.84 (0.01)	&	-		&	-		&	Swope	\\
58704.40	&	-		&	16.64 (0.01)	&	16.82 (0.01)	&	-		&	-		&	LCO	\\
58704.40	&	-		&	16.62 (0.01)	&	16.82 (0.01)	&	-		&	-		&	LCO	\\
58704.48	&	-		&	16.72 (0.01)	&	16.87 (0.02)	&	17.11 (0.04)	&	-		&	Thacher	\\
58705.49	&	-		&	16.53 (0.01)	&	16.72 (0.01)	&	16.93 (0.04)	&	17.06 (0.02)	&	Thacher	\\
58706.10	&	-		&	16.53 (0.01)	&	-		&	16.83 (0.01)	&	-		&	LCO	\\
58706.10	&	-		&	16.55 (0.01)	&	16.76 (0.01)	&	17.03 (0.01)	&	-		&	LCO	\\
58706.40	&	-		&	16.36 (0.01)	&	16.56 (0.01)	&	16.81 (0.01)	&	-		&	LCO	\\
58706.40	&	-		&	16.36 (0.01)	&	16.54 (0.01)	&	16.79 (0.01)	&	-		&	LCO	\\
58706.48	&	-		&	16.42 (0.01)	&	16.57 (0.01)	&	16.82 (0.02)	&	16.98 (0.02)	&	Thacher	\\
58707.30	&	-		&	16.36 (0.01)	&	16.42 (0.01)	&	16.71 (0.01)	&	-		&	LCO	\\
58707.30	&	-		&	16.32 (0.01)	&	16.44 (0.01)	&	16.73 (0.01)	&	-		&	LCO	\\
58707.70	&	-		&	16.30 (0.01)	&	16.45 (0.01)	&	-		&	-		&	LCO	\\
58707.70	&	-		&	16.31 (0.01)	&	16.45 (0.01)	&	16.73 (0.01)	&	-		&	LCO	\\
58710.30	&	-		&	16.40 (0.01)	&	-		&	-		&	-		&	LCO	\\
58710.30	&	-		&	16.34 (0.02)	&	-		&	-		&	-		&	LCO	\\
58710.40	&	-		&	16.30 (0.01)	&	16.44 (0.01)	&	16.70 (0.02)	&	-		&	LCO	\\
58710.40	&	-		&	16.31 (0.01)	&	16.42 (0.01)	&	-		&	-		&	LCO	\\
58710.48	&	-		&	16.35 (0.02)	&	-		&	-		&	-		&	Thacher	\\
58712.40	&	-		&	16.60 (0.01)	&	16.43 (0.01)	&	16.68 (0.01)	&	-		&	Swope	\\
58713.00	&	-		&	16.77 (0.02)	&	16.65 (0.02)	&	16.68 (0.02)	&	-		&	LCO	\\
58713.00	&	-		&	16.75 (0.02)	&	16.59 (0.02)	&	16.84 (0.02)	&	-		&	LCO	\\
58714.10	&	-		&	16.95 (0.01)	&	-		&	-		&	-		&	LCO	\\
58714.10	&	-		&	16.94 (0.01)	&	16.64 (0.01)	&	-		&	-		&	LCO	\\
58714.20	&	-		&	-		&	16.67 (0.01)	&	16.72 (0.02)	&	-		&	LCO	\\
58714.20	&	-		&	-		&	-		&	16.70 (0.01)	&	-		&	LCO	\\
58715.28	&	18.34 (0.03)	&	17.11 (0.01)	&	16.59 (0.01)	&	16.74 (0.01)	&	-		&	Swope	\\
58715.30	&	-		&	17.08 (0.01)	&	16.55 (0.01)	&	16.68 (0.01)	&	-		&	LCO	\\
58715.30	&	-		&	17.09 (0.01)	&	16.56 (0.01)	&	16.68 (0.01)	&	-		&	LCO	\\
58717.29	&	18.91 (0.05)	&	17.45 (0.02)	&	16.70 (0.01)	&	16.74 (0.01)	&	-		&	Swope	\\
58717.40	&	-		&	17.48 (0.01)	&	16.72 (0.01)	&	16.78 (0.01)	&	-		&	LCO	\\
58717.40	&	-		&	17.49 (0.01)	&	16.71 (0.01)	&	16.77 (0.01)	&	-		&	LCO	\\
58717.49	&	-		&	17.40 (0.02)	&	16.72 (0.02)	&	16.82 (0.03)	&	16.89 (0.05)	&	Thacher	\\
58718.40	&	-		&	17.69 (0.01)	&	16.74 (0.01)	&	16.79 (0.01)	&	-		&	LCO	\\
58718.40	&	-		&	17.70 (0.01)	&	16.75 (0.01)	&	16.79 (0.01)	&	-		&	LCO	\\
58719.38	&	19.32 (0.04)	&	17.69 (0.01)	&	16.77 (0.01)	&	16.79 (0.01)	&	-		&	Swope	\\
58720.43	&	-		&	17.82 (0.02)	&	16.88 (0.01)	&	16.85 (0.01)	&	-		&	Swope	\\
58724.40	&	-		&	18.43 (0.01)	&	17.22 (0.01)	&	17.18 (0.01)	&	-		&	LCO	\\
58724.40	&	-		&	18.44 (0.01)	&	17.20 (0.01)	&	17.14 (0.01)	&	-		&	LCO	\\
58725.38	&	20.11 (0.04)	&	18.35 (0.02)	&	17.20 (0.01)	&	17.20 (0.01)	&	-		&	Swope	\\
58726.36	&	20.17 (0.04)	&	18.37 (0.01)	&	17.31 (0.01)	&	17.21 (0.01)	&	-		&	Swope	\\
58727.32	&	20.18 (0.04)	&	18.44 (0.01)	&	17.34 (0.01)	&	17.30 (0.01)	&	-		&	Swope	\\
58727.50	&	-		&	18.44 (0.03)	&	17.45 (0.02)	&	17.36 (0.02)	&	17.40 (0.03)	&	Thacher	\\
58728.10	&	-		&	18.66 (0.02)	&	-		&	-		&	-		&	LCO	\\
58728.10	&	-		&	18.67 (0.02)	&	17.50 (0.01)	&	-		&	-		&	LCO	\\
58728.20	&	-		&	-		&	17.53 (0.01)	&	17.49 (0.02)	&	-		&	LCO	\\
58728.20	&	-		&	-		&	-		&	17.44 (0.02)	&	-		&	LCO	\\
58729.32	&	20.32 (0.04)	&	18.50 (0.01)	&	17.43 (0.01)	&	17.40 (0.01)	&	-		&	Swope	\\
58732.50	&	-		&	18.76 (0.02)	&	17.68 (0.02)	&	-		&	-		&	LCO	\\
58732.50	&	-		&	18.76 (0.02)	&	17.68 (0.02)	&	17.60 (0.02)	&	-		&	LCO	\\
58733.38	&	-		&	18.78 (0.02)	&	17.78 (0.01)	&	17.70 (0.01)	&	-		&	Swope	\\
58736.34	&	-		&	18.78 (0.02)	&	17.75 (0.01)	&	17.68 (0.01)	&	-		&	Swope	\\
58736.60	&	-		&	18.92 (0.03)	&	17.89 (0.02)	&	17.60 (0.02)	&	-		&	LCO	\\
58736.60 	&	-		&	18.88 (0.03)	&	17.89 (0.02)	&	17.83 (0.02)	&	-		&	LCO	\\
58737.40 	&	-		&	18.81 (0.01)	&	17.81 (0.01)	&	17.78 (0.01)	&	-		&	Swope	\\
58740.20 	&	-		&	18.76 (0.11)	&	17.92 (0.05)	&	17.84 (0.06)	&	-		&	LCO	\\
58740.20	&	-		&	19.02 (0.11)	&	18.05 (0.06)	&	17.91 (0.06)	&	-		&	LCO	\\
58744.30 	&	-		&	18.99 (0.04)	&	-		&	-		&	-		&	LCO	\\
58745.32	&	-		&	18.99 (0.03)	&	18.09 (0.02)	&	18.06 (0.02)	&	-		&	Swope	\\
58746.33 	&	-		&	19.02 (0.03)	&	18.16 (0.01)	&	18.13 (0.01)	&	-		&	Swope	\\
58748.39	&	-		&	19.08 (0.01)	&	18.18 (0.01)	&	18.14 (0.01)	&	-		&	Swope	\\
58752.10 	&	-		&	19.15 (0.02)	&	18.29 (0.02)	&	18.24 (0.03)	&	-		&	LCO	\\
58752.10	&	-		&	19.16 (0.01)	&	18.29 (0.02)	&	18.19 (0.03)	&	-		&	LCO	\\
58752.32	&	21.20 (0.07)	&	19.14 (0.01)	&	18.28 (0.01)	&	18.19 (0.01)	&	-		&	Swope	\\
58753.13	&	-		&	19.18 (0.01)	&	13.39 (0.01)	&	18.30 (0.01)	&	-		&	Swope	\\
58754.29	&	21.20 (0.08)	&	19.17 (0.01)	&	18.31 (0.01)	&	18.24 (0.01)	&	-		&	Swope	\\
58755.25 	&	21.09 (0.06) 	&	19.09 (0.01)	&	18.28 (0.01)	&	18.11 (0.01)	&	-		&	Swope	\\
\end{longtable}
\end{center}

\pagebreak

\begin{center}
\begin{longtable}{ccccc} 
\caption{Log of the spectra; $t_\rmn{exp}$ shows the time since the date of explosion constrained in the abundance tomography (MJD $58\,697.5$); while the phases are given relative to the maximum in B-band (MJD $58\,707.8$). } \label{tab:spectroscopic_data} \\
\hline
\multicolumn{1}{c}{\textbf{MJD}} & \multicolumn{1}{c}{\textbf{t$_\rmn{exp}$ [d]}} & \multicolumn{1}{c}{\textbf{Phase [d]}} & \multicolumn{1}{c}{\textbf{Telescope/Instrument}} & \multicolumn{1}{c}{\textbf{Wavelength range [\r{A}]}} \\ \hline 
		58702.8 & 5.3 & -5.0 & LCO/FLOYDS & 3800 - 10000\\
		58703.6 & 6.1 & -4.2 & LCO/FLOYDS & 3500 - 10000\\
		58704.3 & 6.8 & -3.5 & NTT/EFOSC2 & 3300 - 9900\\
		58705.1 & 7.6 & -2.7 & SALT/RSS & 3500 - 9300\\
		58705.4 & 7.9 & -3.4 & Gemini-S/FLAMINGOS-2 & 9900 - 18000\\
		58705.6 & 8.1 & -2.2 & LCO/FLOYDS & 3500 - 9900\\
		58706.1 & 8.6 & -1.7 & SALT/RSS & 3500 - 9300\\
		58707.6 & 10.1 & -0.2 & LCO/FLOYDS & 3500 - 9900\\
		58708.8 & 11.3 & +1.0 & LCO/FLOYDS & 3500 - 9900\\
		58709.8 & 11.3 & +2.0 & LCO/FLOYDS & 3500 - 9900\\
		58717.3 & 19.8 & +9.5 & NTT/EFOSC2 & 3300 - 9900\\
		58718.2 & 20.7 & +10.4 & NTT/EFOSC2 & 3600 - 9200\\
		58718.8 & 21.3 & +11.0 & LCO/FLOYDS & 3800 - 9900\\
		58721.5 & 24.0 & +13.7 & Lick/Kast & 3500 - 10500\\
		58724.2 & 26.7 & +16.6 & NTT/EFOSC2 & 3300 - 7400\\
		58725.0 & 27.5 & +17.2 & SALT/RSS & 3500 - 9300\\
		58727.5 & 30.00 & +19.7 & Lick/Kast & 3500 - 10500\\
		58727.6 & 30.1 & +19.8 & LCO/FLOYDS & 3800 - 9900\\
		58730.2 & 32.7 & +22.4 & NTT/EFOSC2 & 6000 - 9900\\
		58734.3 & 36.8 & +26.5 & Gemini-S/FLAMINGOS-2 & 9900 - 18000\\
		58735.5 & 38.0 & +27.7 & Lick/Kast & 3500 - 10500\\
		58736.6 & 39.1 & +28.8 & LCO/FLOYDS & 3800 - 9900\\
		58752.9 & 55.4 & +45.1 & SALT/RSS & 3500 - 9300\\
		58757.9 & 60.4 & +50.1 & SALT/RSS & 3500 - 9300\\
		\hline
		\hline
\end{longtable}
\end{center}

\begin{center}
\begin{longtable}{ccccccc}
\caption{Optical light curve parameters of SN 2019muj. The rising times ($t_\rmn{rise}$) are based on the date of explosion estimated from the quasi-bolometric light curve fitting.} \label{tab:photometric_fitting} \\
\hline
\multicolumn{1}{c}{\textbf{Filter}} & \multicolumn{1}{c}{\textbf{U}} & \multicolumn{1}{c}{\textbf{B}} & \multicolumn{1}{c}{\textbf{g}} & \multicolumn{1}{c}{\textbf{V}} & \multicolumn{1}{c}{\textbf{r}} & \multicolumn{1}{c}{\textbf{i}}\\
\hline
		 t$_\rmn{max}$ & 58706.7(0.5)  & 58707.8 (0.5) & 58708.2(0.6) & 58709.3(0.7) & 58709.7(0.9) & 58710.5(1.2) \\
		 Peak Obs. [mag] & 15.88(0.08)  & 16.42(0.05) & 16.29(0.06) & 16.33(0.05) & 16.39(0.07) & 16.61(0.07) \\
		 Peak Abs. [mag] & $-$16.92(0.09)  & $-$16.36(0.06) & $-$16.48(0.07) & $-$16.42(0.06) & $-$16.35(0.08) & $-$16.12(0.09) \\
		 $\Delta$m$_{15}$ [mag] & 3.1 & 2.4 & 2.0 & 1.2 & 1.0 & 0.7 \\
		 t$_\rmn{rise}$ [days] & 8.5(0.7) & 9.6(0.7) & 10.0(0.8) & 10.5(0.8) & 11.5(1.1) & 12.3(1.3) \\
\hline
\hline
\end{longtable}
\end{center}

\begin{center}
\begin{longtable}{cccc}
\caption{Fitting parameters of the quasi-bolometric light curve of SN 2019muj described in Sec. \ref{bolometric}.} \label{tab:bolometric_fitting} \\
\hline
\multicolumn{1}{c}{\textbf{M$_{^{56}Ni}$ [ M$_\odot$]}} & \multicolumn{1}{c}{\textbf{t$_\rmn{exp} [MJD]$}} & \multicolumn{1}{c}{\textbf{t$_\rmn{d}$ [days]}} & \multicolumn{1}{c}{\textbf{A$_\gamma$}}\\
\hline
		 0.031 $\pm$ 0.005 & 58698.2 $\pm$ 0.5 & 7.1 $\pm$ 0.5  & 5.5 $\pm$ 1.5\\
		\hline
		\hline
\end{longtable}
\end{center}


\bsp	
\label{lastpage}
\end{document}